\def\year{2018}\relax
\documentclass[letterpaper]{article} 

\usepackage{amsmath}
\usepackage{amsthm}
\usepackage{amsfonts}
\usepackage{amssymb}
\usepackage[linesnumbered,ruled]{algorithm2e}
\usepackage{subcaption}
\usepackage{color}

\usepackage{aaai18}  
\usepackage{times}  
\usepackage{helvet}  
\usepackage{courier}  
\usepackage{url}  
\usepackage{graphicx}  
\begin{document}
%
\title{Geographic Differential Privacy for Mobile Crowd Coverage Maximization}
\author{Leye Wang\textsuperscript{1}, Gehua Qin\textsuperscript{2}, Dingqi Yang\textsuperscript{3}, Xiao Han\textsuperscript{4}, Xiaojuan Ma\textsuperscript{1}\\
\textsuperscript{1}The Hong Kong University of Science and Technology,
\textsuperscript{2}Shanghai Jiao Tong University\\
\textsuperscript{3}University of Fribourg,
\textsuperscript{4}Shanghai University of Finance and Economics\\
wly@cse.ust.hk, qingehua@gmail.com, dingqi@exascale.info, xiaohan@mail.shufe.edu.cn, mxj@cse.ust.hk
}

\maketitle
\begin{abstract}
For real-world mobile applications such as location-based advertising and spatial crowdsourcing, a key to success is targeting mobile users that can maximally cover certain locations in a future period. To find an optimal group of users, existing methods often require information about users' mobility history, which may cause privacy breaches. In this paper, we propose a method to maximize mobile crowd's future location coverage  under a guaranteed location privacy protection scheme. In our approach, users only need to upload one of their frequently visited locations, and more importantly, the uploaded location is obfuscated using a geographic differential privacy policy. We propose both analytic and practical solutions to this problem. Experiments on real user mobility datasets show that our method significantly outperforms the  state-of-the-art geographic differential privacy methods by achieving a higher coverage under the same level of privacy protection.
\end{abstract}

\section{Introduction} 
\label{sec:introduction}

Crowd coverage maximization is a classical problem in mobile computing: how to select $m$ users from a candidate pool to maximize the probability of covering a set of target locations in a coming time period (e.g., one day or one week). 
This problem and its variants have a wide spectrum of applications in location-based advertising~\cite{dhar2011challenges}, spatial crowdsoucing~\cite{chen2016spatial,zhang20144w1h}, urban computing~\cite{zheng2014urban}, etc. For example, it can help shop owners to offer electronic coupons to the set of mobile app users who may physically visit the region around the shop soon; it can also help crowdsourcing organizers to recruit the participants to cover the task area with the highest probability~\cite{xiong2016icrowd}.

One of the key steps in crowd coverage maximization is \textit{mobility profiling}, i.e., predicting the probability of a user appearing at a certain location. A common practice is first dividing an area into fine-grained grids or sub-areas, and then counting the frequency of a user appearing in each grid based on trajectory history~\cite{guo2017activecrowd}. One can use more sophisticated models like Poisson process to estimate users' occurrence distribution~\cite{xiong2016icrowd}. Existing mobility profiling  methods often require access to users' historical mobility traces, which may seriously compromise user privacy. For example, 
users' exposed location data may reveal sensitive information about their identities and social relationships~\cite{cho2011friendship,rossi2015privacy}. Despite the importance of location privacy, as far as we know, there is little research effort combining location privacy, mobility profiling, and crowd coverage maximization up to date.

To fill this gap, this paper aims to explore how to protect the crowds' location privacy, while still optimizing their expected coverage of a set of locations. To achieve this goal, we propose a mobile crowd coverage maximization framework with a rigorous privacy protection scheme --- \textit{geographic differential privacy}~\cite{andres2013geo}. A geographic differential privacy policy obfuscates a user's actual location to another with carefully designed probabilities, such that adversaries, regardless of their prior knowledge, can learn little about the user's true location after observing the obfuscated locations. However, with differential privacy protection, crowd coverage maximization can only be performed based on the obfuscated (inaccurate) locations, which leads to inevitable loss of the quality of the selected users. Therefore, we propose a method to generate the optimal location obfuscation policy which satisfies geographic differential privacy while minimizing such loss. 

In summary, this paper has the following contributions:

(1) To the best of our knowledge, this is the first work studying the mobile crowd coverage maximization problem with location privacy protection.

(2) In our approach, users only need to upload one of their frequently visited locations, and more importantly, the uploaded location is obfuscated using the rigorous privacy policy --- geographic differential privacy. We further formulate an optimization problem to obtain the optimal obfuscation policy that can maximize the expected future crowd coverage over a set of locations under a guaranteed level of differential privacy protection. As the optimization problem is non-convex, we first mathematically analyze the scenario when only one location needs to cover and then derive an optimal solution. Then, we extend this setting to the multi-location coverage scenario and propose a practical algorithm to obtain the optimal obfuscation policy.

(3) Experiments on real human mobility datasets verify that, by selecting the same number of users under the same level of privacy protection, our method achieves a higher coverage than state-of-the-art differential privacy methods. 

\section{Preliminaries} 

\textbf{Geographic differential privacy}~\cite{andres2013geo} introduces the idea of database differential privacy~\cite{dwork2008differential} into the location obfuscation context. Its key idea is: given an observed obfuscated location $l^*$, any two locations $l_1$ and $l_2$ have similar probabilities of being mapped to $l^*$. It is thus hard for an adversary to differentiate whether the user is at $l_1$ or $l_2$ by observing $l^*$.

\textbf{Definition 1}~\cite{andres2013geo}. \textit{Suppose the target area includes a set of locations $\mathcal{L}$, then an obfuscation policy $P$ satisfies \textbf{geographic $\epsilon$-differential privacy}, iff. }
\begin{equation}
P(l^*|l_1) \leq e^{\epsilon d(l_1,l_2)} P(l^*|l_2) \quad\quad \forall l_1,l_2,l^* \in \mathcal{L}
\label{eq:dp}
\end{equation}
\textit{where $P(l^*|l)$ is the probability of obfuscating $l$ to $l^*$, $d(l_1,l_2)$ is the distance between $l_1$ and $l_2$, $\epsilon$ is the privacy budget --- the smaller $\epsilon$, the better privacy protection.}

Note that the set of locations are usually constructed by dividing the target area into subregions, e.g., equal-size grids~\cite{bordenabe2014optimal} or cell-tower regions~\cite{xiong2016icrowd}. 

If $P$ satisfies geographic differential privacy, it can be proven that for adversaries with \textit{any} prior knowledge about users' location distributions, their posterior knowledge after observing the obfuscated location can only be increased by a small constant factor \cite{andres2013geo}. Note that this protection is guaranteed even if the adversaries know $P$. Due to this rigorous protection effect, geographic differential privacy has seen many applications in location based services, spatial crowdsourcing, etc. \cite{bordenabe2014optimal,wang2016differential,wang2017location}.

\textbf{Mobility profiling} aims to estimate the probability of a user covering a certain location during a time period in the future. Specifically, a user $u_i$'s mobility profile is denoted as $M_i$, and $M_i(l_j), \ l_j \in \mathcal L$ means the estimated probability of $u_i$ visiting $l_j$ in a concerned future period (e.g., next week). Commonly used mobility profiling methods include frequency-based~\cite{guo2017activecrowd} and Poisson-based~\cite{xiong2016icrowd} algorithms. We use the Poisson process to model user mobility given its better prediction performance in our experiments. More details can be found in the appendix.

\section{Framework Overview} 

\begin{figure}[t]
	\centering
	\includegraphics[width=1\linewidth]{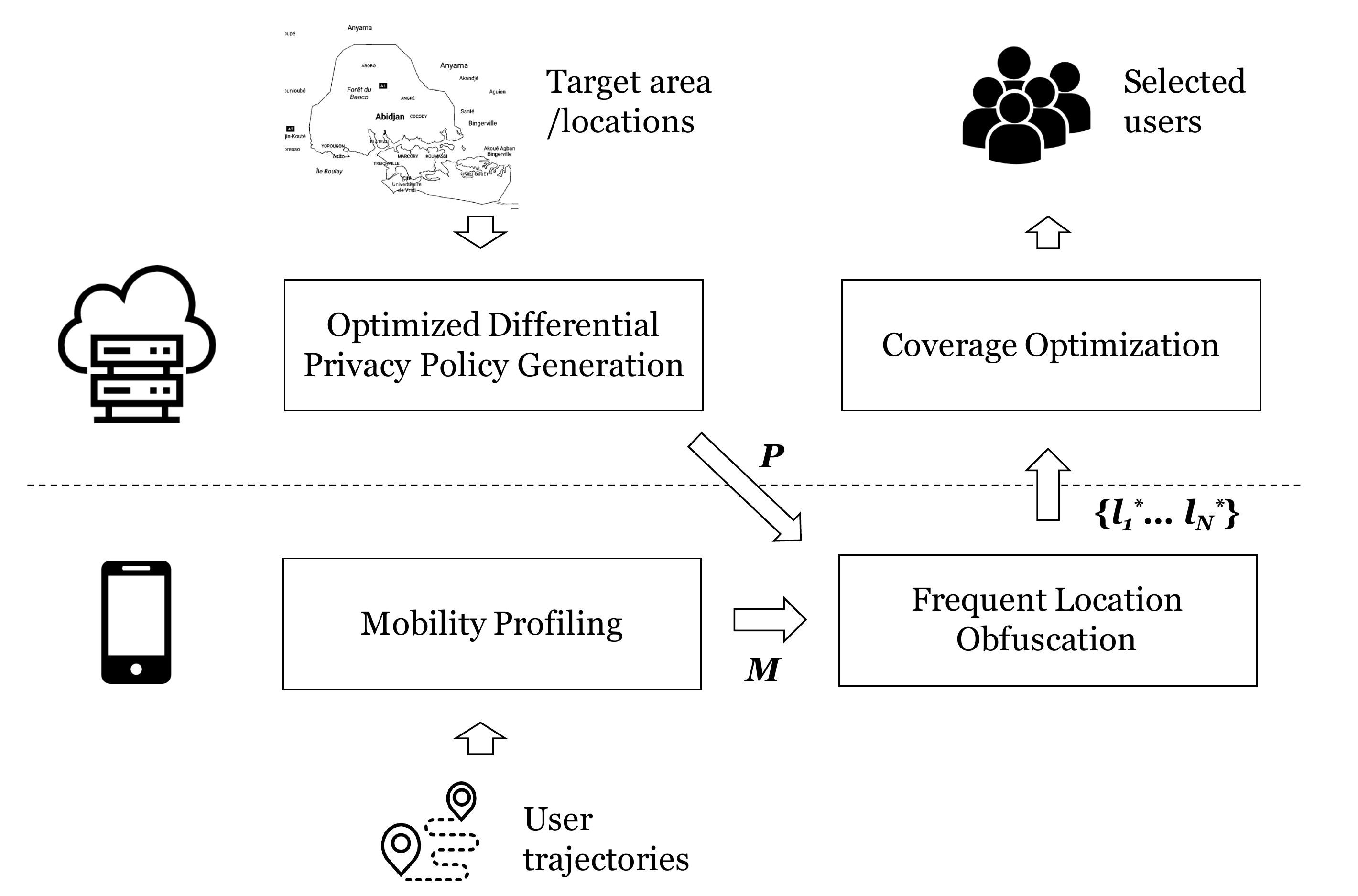}
	\caption{Framework overview.}
	\label{fig:framework}
\end{figure}

We present an overview of our privacy framework in Figure~\ref{fig:framework}. The key idea of our framework is that users should expose their location information as little as possible, while we can still select a proper set of users for optimizing their coverage on certain target locations in the future.

The two main players in our framework are a server platform and its mobile client users. As we want users to expose their actual location information as little as possible, user mobility profiling runs locally on individuals' smart devices. That means, the clients' mobility profiles are only known to themselves. As shown in the literature, only uploading frequent locations with high profiling probabilities (e.g., $> 80\%$) to the server can already help achieve a good future crowd coverage~\cite{guo2017activecrowd}. To limit the potential location leakage, our framework only requires users to upload \textit{one} of their frequent locations. Moreover, this frequent location is obfuscated by the geographic differential privacy policy $P$ before being sent to the platform. The policy $P$ is generated by the server based on which target locations need to be covered. Finally, according to the uploaded obfuscated frequent locations $\{l_1^*\cdots l_N^*\}$ (suppose $N$ users), the platform aims to select a set of users to maximize the expected coverage of intended locations in the coming period.

\begin{figure}[t]
	\centering
	\includegraphics[width=1\linewidth]{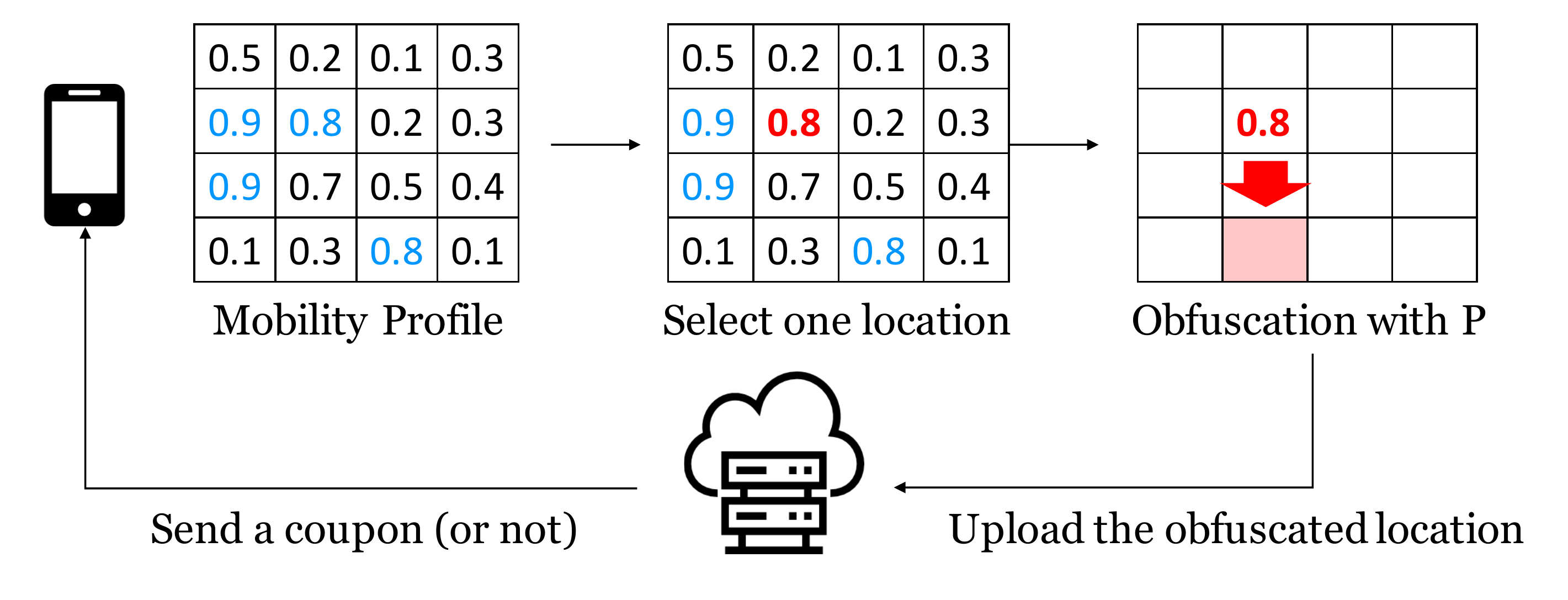}
	\caption{A running example of our framework.}
	\label{fig:run_example}
\end{figure}

A running example is shown in Figure~\ref{fig:run_example}, where the table represents a user's mobility profile in a 2D spatial area splitted into uniform grids. Suppose that a location-based advertising platform needs to decide whether to send a Starbucks coupon to a user. The platform expects that a user receiving the coupon is a frequent visitor to the regions where Starbucks stores are located, so that the user will probably go to the stores. To achieve this goal, first, a user client computes its owner's mobility profile locally. Second, from the set of locations whose probabilities are larger than a threshold (e.g., 80\%), the user client randomly selects one location $l_u$ to be uploaded to the server.\footnote{If there is no location with the probability larger than the threshold, then the user does not upload any location.} Third, according to the privacy policy $P$ received from the server, the user client randomly obfuscates $l_u$ to $l_u^*$ and then sends $l_u^*$ to the server. Finally, the server will decide whether to send the coupon to the user or not according to  the uploaded $l_u^*$. In this case, a user's location privacy is preserved as the uploaded frequent location is rigorously obfuscated with differential privacy.

Location obfuscation would inevitably introduce certain loss of quality in selecting users for coverage optimization, as users' uploaded frequent locations contain deliberate noises. Hence, how the server generates the privacy policy $P$ is the key challenge of our framework, which aims to minimize the loss of quality caused by privacy protection.

\section{Optimal Privacy Policy} 
\label{sec:optimal_privacy_policy}

In this section, we illustrate our solution that guarantees geographic $\epsilon$-differential privacy while minimizing the loss of quality in mobile crowd coverage optimization.

\subsection{Single Location Coverage Problem (SLCP)}
\label{sub:one_location}

As the first step, we analyze the scenario where only one location needs to be covered. In location-based advertising, this reflects the scenario that the advertising only involves one specific site (e.g., a newly opened restaurant). In spatial crowdsourcing, this means that the task is only associated with one location (e.g., taking the photo of Statue of Liberty). Suppose the target location to cover as $l_t$ and a user submits her/his obfuscated frequent location as $l^*$, then the probability of her/his frequent location being actually $l_t$ is:
\begin{equation}
prob(l_t|l^*) = \frac{\pi(l_t)P(l^*|l_t)}{\sum_{l \in \mathcal L} \pi(l) P(l^*|l)} 
\label{eq:one_location}
\end{equation} 
where $\pi$ is the overall distribution of all the users' frequent locations. Here we suppose that we can foreknow $\pi$, and later we will elaborate how to estimate it.  Note that the denominator can be seen as the overall probability of a user reporting her/his frequent location as $l^*$.

Suppose we select a user reporting $\hat l^*$ to cover the target location $l_t$ in the coming time period, apparently we would like to maximize Eq.~\ref{eq:one_location} so that the future probability of the user covering $l_t$ is maximized. With this idea, we have the following optimization process to get the optimal privacy policy $\hat P$. Particularly, given $l_t$ to cover, we aim to
\begin{align}
	& \max_{\hat l^*,\hat P} \frac{\pi(l_t)\hat P(\hat l^*|l_t) }{\sum_{l \in \mathcal L} \pi(l) \hat P(\hat l^*|l)}
	\label{eq:max_one_task_location}
	\\
	s.t. \quad & \hat P(l^*|l_1) \leq e^{\epsilon d(l_1,l_2)} \hat P(l^*|l_2)  && \forall l_1,l_2,l^* \in \mathcal{L} 
	\label{eq:const_start}
	\\
	& \hat P(l^*|l) > 0  &&\forall l, l^* \in \mathcal L 
	\label{eq:const_start2}
	\\
	& \sum_{l^* \in \mathcal L} \hat P(l^*|l) = 1 &&\forall l \in \mathcal L 
	\label{eq:const_dp_end}
\end{align}
Eq.~\ref{eq:const_start} is the constraint of geographic differential privacy; Eq.~\ref{eq:const_start2} and \ref{eq:const_dp_end} are probability restrictions.
By solving the above optimization problem, we can get the optimal privacy policy $\hat P$, as well as the user selection strategy, i.e., selecting the users reporting $\hat l^*$ for future coverage maximization. 

However, even given $\hat l^*$, Eq.~\ref{eq:max_one_task_location} cannot be converted to a convex optimization problem with existing solutions \cite{boyd2004convex}. To overcome this difficulty, we then analyze the relationship between the constraints and the objective function, and then deduce an optimal solution analytically.

\subsection{An Analytic Solution to SLCP}

Our analytic deduction includes three steps. First, we verify that the selection of $\hat l^*$ will not affect the optimal objective value of Eq.~\ref{eq:max_one_task_location}. Second, we prove that Eq.~\ref{eq:max_one_task_location} cannot exceed a certain upper bound. Finally, we show that this upper bound can be achieved by constructing a feasible solution of $\hat P$.

\textbf{Lemma 1.} \textit{For any two locations $l_1^*, l_2^* \in \mathcal L$, the optimal objective values of Eq.~\ref{eq:max_one_task_location} are the same if we set $\hat l^* = l_1^*$ or $l_2^*$.}

\textit{Proof.} For $\hat l^* = l_1^*$ or $l_2^*$, we can always find a pair of $P_1$, $P_2$, where $P_1(l_1^*|l)= P_2(l_2^*|l), P_1(l_2^*|l)=P_2(l_1^*|l)$, and $P_1(l^*|l)=P_2(l^*|l)$ for other $l^*$; $P_1$ and $P_2$ lead to the same objective value. A detailed proof is in the appendix. \qedsymbol

\textit{Remark.} Lemma 1 demonstrates that we can use any location $l \in \mathcal L$ as the obfuscated location $\hat l^*$ for user selection without impacting the achievable optimal coverage utility.

\textbf{Lemma 2.} \textit{The optimal value of Eq.~\ref{eq:max_one_task_location} cannot exceed}
\begin{equation}
\frac{\pi (l_t)}{\sum_{l \in \mathcal L} \pi(l) e^{-\epsilon d(l,l_t)}}
\label{eq:max_obj_value}
\end{equation}
\textit{and this value can only be achieved if }
\begin{equation}
\hat P(\hat l^*|l) \propto e^{-\epsilon d(l,l_t)}, \quad \forall l \in \mathcal L
\label{eq:p_max}
\end{equation}

\textit{Proof.} With geographic differential privacy constraints, 
\begin{align}
	&\frac{\pi(l_t)\hat P(\hat l^*|l_t) }{\sum_{l \in \mathcal L} \pi(l) \hat P(\hat l^*|l)} = \frac{\pi(l_t)}{\sum_{l \in \mathcal L} \pi(l) \frac{\hat P(\hat l^*|l)}{\hat P(\hat l^*|l_t)}} \\
	\le &\frac{\pi (l_t)}{\sum_{l \in \mathcal L} \pi(l) \frac{e^{-\epsilon d(l,l_t)} \hat P(\hat l^*|l_t)}{\hat P(\hat l^*|l_t)}} =  \frac{\pi (l_t)}{\sum_{l \in \mathcal L} \pi(l) e^{-\epsilon d(l,l_t)}}
\end{align}

\qedsymbol

\textit{Remark.} Lemma 2 points out an upper bound of the optimal objective value and the condition (Eq.~\ref{eq:p_max}) that $\hat P$ must satisfy for getting the upper bound value. However, whether we can find a feasible $\hat P$ satisfying Eq.~\ref{eq:p_max}, as well as Eq.~\ref{eq:const_start}-\ref{eq:const_dp_end} is still unknown. Next, we prove that this $\hat P$ exists.

\textbf{Lemma 3.} \textit{If $\hat P(\hat l^*|l)$ satisfies Eq.~\ref{eq:p_max}, then}
\begin{equation}
\hat P(\hat l^*|l_1) \leq e^{\epsilon d(l_1,l_2)} \hat P(\hat l^*|l_2) \quad \forall l_1, l_2 \in \mathcal{L} 
\end{equation}

\textit{Proof. } Considering that $d$ is a distance metric, then
\begin{equation}
\frac{\hat P(\hat l^*|l_1)}{\hat P(\hat l^*|l_2)} = e^{\epsilon (d(l_2, l_t) - d(l_1, l_t))} \le e^{\epsilon d(l_1, l_2)}
\end{equation}

\qedsymbol

\textit{Remark.} Lemma 3 proves that when Eq.~\ref{eq:p_max} stands, Eq.~\ref{eq:const_start} of $l^* = \hat l^*$ must also hold for any $l_1$, $l_2$.

\textbf{Theorem 1.} Given any $\hat l^*$, we can get a feasible $\hat P$,
\begin{align}
	\hat P(\hat l^*|l) & = \theta e^{-\epsilon d(l, l_t)}, && \forall l \in \mathcal L 
	\label{eq:optimal_p_1}
	\\
	\hat P(l^*|l) & = \frac{1-\theta e^{-\epsilon d(l, l_t)}}{|\mathcal L|-1}, && \forall  l^*, l \in \mathcal L \textit{ and } l^* \not = \hat l^*
	\label{eq:optimal_p_2}
\end{align}
which can achieve the upper bound Eq.~\ref{eq:max_obj_value}. Here, $\theta$ can be any positive constant value smaller than or equal to a threshold $\tau$, where
\begin{equation}
\tau =  \min_{l_1,l_2 \in \mathcal L} \frac{e^{\epsilon d(l_1, l_2)}-1}{e^{-\epsilon (d(l_2,l_t)-d(l_1,l_2))} - e^{-\epsilon d(l_1,l_t)}}
\end{equation}

The proof is in the appendix.

Note that while Theorem~1 gets an optimal solution, in reality, there may not be enough users who report $\hat l^*$ for selection (if $\tau$ is too small and the total user number is limited). Later we will propose a practical solution overcoming this shortcoming, when addressing the multi-location scenario.

\subsection{Multi-Location Coverage Problem (MLCP)}
\label{sub:multi_loc}

A more complicated setting for mobile crowd coverage problem includes a set of locations that need to be covered. Real-life examples include delivering coupons of chain stores to users who will probably visit any of them in the next time period.
Denote the set of locations to cover as 
\begin{equation}
\mathbb L = \{l_t^1,l_t^2,...,l_t^z\} \subset \mathcal L
\end{equation}
then the probability of a user's actual frequent location belonging to $\mathbb L$ is:
\begin{equation}
\sum_{l_t \in \mathbb L} prob(l_t|l^*) =  \frac{\sum_{l_t \in \mathbb L} \pi(l_t)P(l^*|l_t)}{\sum_{l \in \mathcal L} \pi(l) P(l^*|l)} 
\label{eq:multi_locations}
\end{equation}

Then, we can maximize Eq.~\ref{eq:multi_locations} with the constraints Eq.~\ref{eq:const_start}-\ref{eq:const_dp_end} to get the optimal privacy policy $\hat P$, and the obfuscated location $\hat l^*$ for future crowd coverage maximization.
\begin{align}
	& \max_{\hat l^*, \hat P} \frac{\sum_{l_t \in \mathbb L} \pi(l_t) \hat P(\hat l^*|l_t)}{\sum_{l \in \mathcal L} \pi(l) \hat P(\hat l^*|l)} 
	\label{eq:max_multi_task_location}
	\\
	& s.t. \quad Eq.~\ref{eq:const_start} - \ref{eq:const_dp_end}
\end{align}

Similar to the single location coverage problem, we can prove the following lemmas.

\textbf{Lemma 4.} \textit{For any $l_1^*, l_2^* \in \mathcal L$, the optimal objective values of Eq.~\ref{eq:max_multi_task_location} are the same if we set $\hat l^* = l_1^*$ or $l_2^*$.}

\textit{Lemma 4} is a straightforward extension of \textit{Lemma 1} to the multiple location coverage scenario.

\textbf{Lemma 5.} \textit{The optimal value of Eq.~\ref{eq:max_multi_task_location} cannot exceed}
\begin{equation}
(1+\sum_{l \not \in \mathbb L}  \sum_{l_t \in \mathbb L} \frac{\pi(l)}{ \pi(l_t)}e^{-\epsilon d(l,l_t)})^{-1}
\label{eq:optimal_obj_multi}
\end{equation}
and this value can be achieved only if
\begin{equation}
e^{\epsilon d(l,l_t)} P(\hat l^*|l) =  P(\hat l^*|l_t), \ \forall l_t \in \mathbb L, \forall l \not \in \mathbb L
\label{eq:optimal_obj_multi_cond}
\end{equation}

The detailed proof is in the appendix.

\begin{figure}[t]
	\centering
	\includegraphics[width=.4\linewidth]{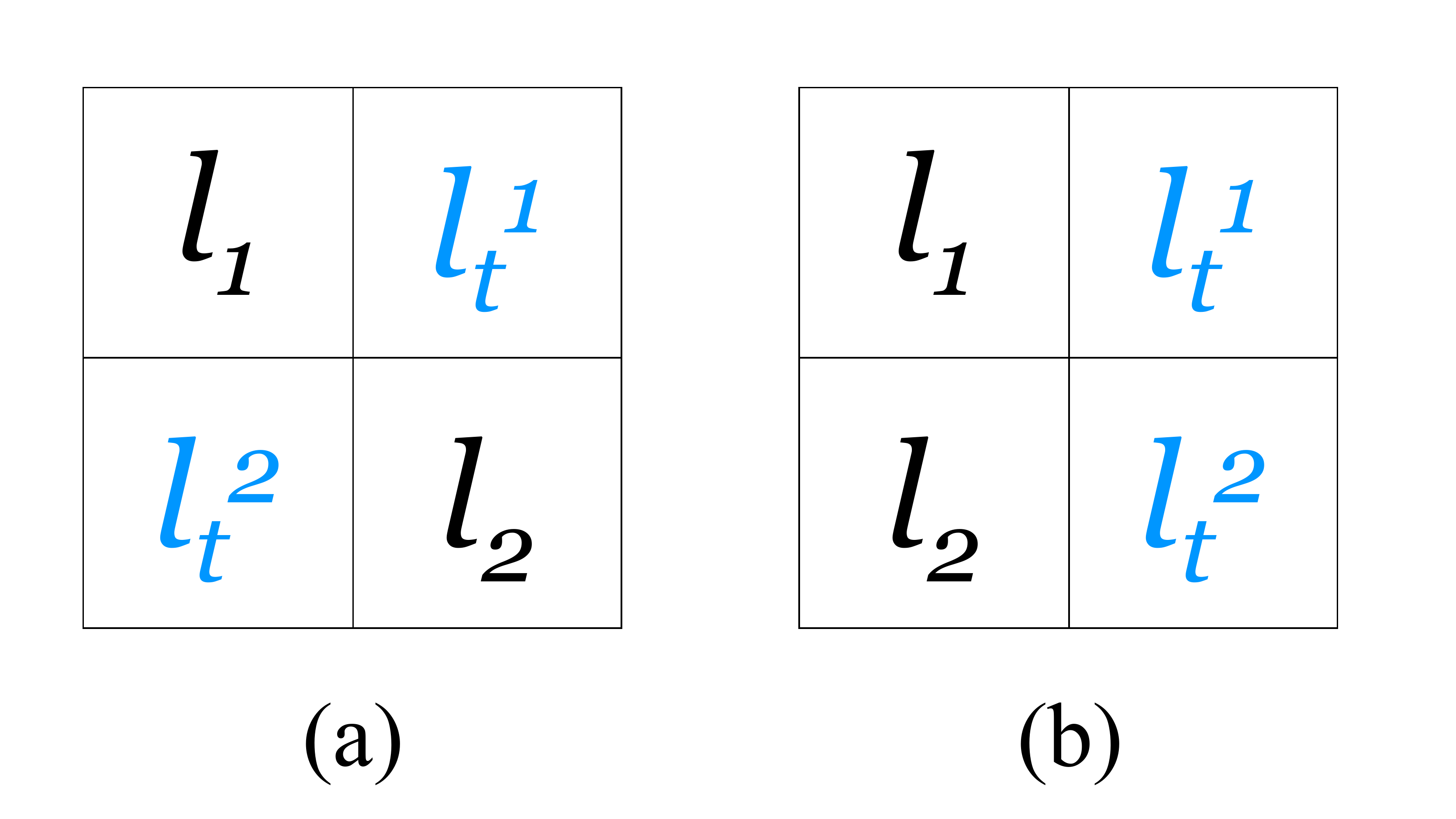}
	\caption{Toy examples with two locations to cover.}
	\label{fig:location_examples}
\end{figure}

Although Lemma 5 seems to be an extension of Lemma 2 for the multi-location scenario, they have a significant difference that the optimal value Eq.~\ref{eq:optimal_obj_multi} may not always be feasible, i.e., Eq.~\ref{eq:optimal_obj_multi_cond} may not stand.
Take a toy example of $\mathbb L$ containing two locations, it means that, for any $l \not \in \mathbb L$
\begin{align}
	e^{\epsilon d(l,l_t^1)} P(\hat l^*|l) =  P(\hat l^*|l_t^1) \\
	e^{\epsilon d(l,l_t^2)} P(\hat l^*|l) =  P(\hat l^*|l_t^2)
\end{align}
Then, for any two locations $l_1, l_2 \not \in \mathbb L$, let $l = l_1$ or $l_2$, then
\begin{align}
	& \frac{P(\hat l^*|l_t^1)}{P(\hat l^*|l_t^2)} = 
	\frac{e^{\epsilon d(l_1,l_t^1)}}{e^{\epsilon d(l_1,l_t^2)}} = \frac{e^{\epsilon d(l_2,l_t^1)}}{e^{\epsilon d(l_2,l_t^2)}} \\
	\Rightarrow \ & d(l_1,l_t^1) - d(l_1,l_t^2) = d(l_2,l_t^1) - d(l_2,l_t^2)
	\label{eq:opt_multi_cond}
\end{align}
Hence, if Eq.~\ref{eq:optimal_obj_multi} is feasible, Eq.~\ref{eq:opt_multi_cond} must hold. Figure~\ref{fig:location_examples} shows two examples, in one of which Eq.~\ref{eq:opt_multi_cond} stands (Figure~\ref{fig:location_examples}a) and the other does not (Figure~\ref{fig:location_examples}b, considering the Euclidean distance). This shows that whether Eq.~\ref{eq:optimal_obj_multi} can be achieved depends on the distribution of the target locations.

\subsection{A Practical Solution to MLCP}

While we cannot always obtain the upper bound value of  Eq.~\ref{eq:optimal_obj_multi} for the multi-location coverage problem, here we propose a practical solution which can work in real scenarios.

Revisiting the objective function of the multi-location coverage problem, Eq.~\ref{eq:max_multi_task_location}, we can see that the main difficulty in solving the optimization problem is that the denominator includes $\hat P$ in it. To address this issue, we propose to add one more constraint to the optimization process by setting the denominator to a constant value,
\begin{equation}
\sum_{l \in \mathcal L} \pi(l) \hat P(\hat l^*|l) = \beta
\label{eq:const_beta}
\end{equation}
where $\beta$ is a constant between 0 and 1; we will later elaborate how to set $\beta$. With Eq.~\ref{eq:const_beta}, the objective function is,
\begin{equation}
\max_{\hat l^*, \hat P} \frac{\sum_{l_t \in \mathbb L} \pi(l_t) \hat P(\hat l^*|l_t)}{\beta} 
\label{eq:max_multi_beta}
\end{equation}
Lemma 4 has shown that we can set $\hat l^*$ to any $l \in \mathcal L$ without affecting the optimal objective value. Since Eq.~\ref{eq:const_start}-\ref{eq:const_dp_end} are all linear constraints, we can then use state-of-the-art linear programming tools (e.g., Mosek and Gurobi) to solve the optimization problem to get the optimal privacy policy $\hat P$.

\subsubsection{Setting $\beta$ with Binomial Distribution.}

We then discuss how to set $\beta$ in real-life scenarios. First, we prove that if we want to get the objective value as high as possible, we should set $\beta$ as small as possible.

\textbf{Theorem 2.} \textit{Given $\hat l^*$, suppose $v_1$, $v_2$ are the two optimal objective values of Eq.~\ref{eq:max_multi_beta} when we set $\beta$ to $b_1$, $b_2$, respectively, and $b_1 < b_2$, then $v_1 \ge v_2$.}

\textit{Proof.} We denote the optimal $\hat P$ when setting $\beta$ to $b_1$, $b_2$ as $\hat P_1$, $\hat P_2$, respectively. Then, we construct a new solution of $P_1'$ when $\beta = b_1$ as follows:
\begin{align*}
	& P_1'(\hat l^*|l) = \theta \hat P_2(\hat l^*|l) &&\forall l \in \mathcal L \\
	& P_1'(l'|l) = \hat P_2(l'|l) + (1-\theta) \hat P_2(\hat l'|l)  &&\forall l \in \mathcal L, l' \not = \hat l^*
\end{align*}
where $\theta = b1/b2$. All the constraints of Eq.~\ref{eq:const_start}-\ref{eq:const_dp_end} still stand for $P_1'$. As the optimal objective value  is $v_1$ when $\beta = b_1$,
\begin{align*}
	v_1 \ge \frac{\sum_{l_t \in \mathbb L} \pi(l_t) P_1'(\hat l^*|l_t)}{b_1}
	= \frac{\sum_{l_t \in \mathbb L} \pi(l_t) \hat P_2(\hat l^*|l_t)}{b_2} = v_2 
\end{align*}

\qedsymbol

\textit{Theorem 2} is very important for our practical solution, because it tells us that to get the optimal solution, we only need to solve the linear program \textit{once} by setting $\beta$ to the smallest value that we can accept, rather than enumerating all the possible $\beta$. On the other hand, $\beta$ can be seen as the overall probability that a user will report her/his frequent location as $\hat l^*$. Since we need to select users from such users, we cannot set $\beta$ to a too small value, which will lead to very few people reporting their locations as $\hat l^*$. Therefore, we propose a method to set $\beta$, with a guarantee that the platform can find $\alpha$ users with a probability of $\rho$ (e.g., 95\%) as follows.

\setlength{\textfloatsep}{3pt}
\begin{algorithm}[t]
	\scriptsize
	\SetKwInOut{Input}{Input}
	\SetKwInOut{Output}{Output}
	\Input{$\pi$: overall user spatial distribution.\\
		$\epsilon$: differential privacy budget.\\
		$\mathcal L$: whole set of locations.\\
		$\mathbb L$: set of target locations to cover.\\
		$N$: total number of users.\\
		$\alpha$: number of users to select.\\
		$\rho$: probability threshold for user selection.
	}
	\Output{$\hat P$: optimal differential privacy policy.\\
		$\hat l^*$: the obfuscated location to select users.}
	$\hat l^* \leftarrow l_1$ (or any other $l \in \mathcal L$) \;
	$\beta \leftarrow$ the minimum value that can ensure $Pr(X \ge \alpha) \ge \rho$ for the Binomial distribution $B(X, N, \beta)$\;
	Solve the linear program to get optimal $\hat P$:
	\begin{align*}
		& \max_{ \hat P} \frac{\sum_{l_t \in \mathbb L} \pi(l_t) \hat P(\hat l^*|l_t)}{\beta} 
		\\
		s.t. \quad & \hat P(l^*|l_1) \leq e^{\epsilon d(l_1,l_2)} \hat P(l^*|l_2)  && \forall l_1,l_2,l^* \in \mathcal{L} 
		\\
		& \hat P(l^*|l) > 0  &&\forall l, l^* \in \mathcal L 
		\\
		& \sum_{l^* \in \mathcal L} \hat P(l^*|l) = 1 &&\forall l \in \mathcal L \\
		& \sum_{l \in \mathcal L} \pi(l) \hat P(\hat l^*|l) = \beta
	\end{align*}\
	
	\Return $\hat l^*$, $\hat P$ \;
	\caption{\footnotesize Optimal policy for multi-location coverage.}
	\label{alg:practical_solution}
\end{algorithm}

Suppose totally $N$ users report their frequent locations, then we can estimate the number of users who will report their obfuscated frequent locations as $\hat l^*$ with the Binomial probability $Pr(X=m) = B(m,N,\beta)$. Then, the probability that we can find at least $\alpha$ users is that,
\begin{equation}
Pr(X\ge \alpha) = \sum_{m=\alpha}^{N} B(m,N,\beta)
\end{equation}
And thus we would like to set $\beta$ to the smallest value that ensures $Pr(X \ge \alpha) \ge \rho$.

We describe the pseudo-code of our practical solution for the private multi-location coverage problem in Algorithm~\ref{alg:practical_solution}. Note that since covering one location is a special case of covering multiple locations, Algorithm~\ref{alg:practical_solution} can also solve the single location coverage problem, without the need to assume that we will always have enough users reporting $\hat l^*$.

\begin{algorithm}[t]
	\scriptsize
	\SetKwInOut{Input}{Input}
	\SetKwInOut{Output}{Output}
	\Input{$k$: number of user groups to split.\\
		other inputs like Algorithm~\ref{alg:practical_solution}, except that $\pi$ is unknown.}
	\Output{$U^*$: selected users.}
	$\pi \leftarrow$ uniform distribution (or other proper initial distribution)\;
	$U_1, U_2, ..., U_k \leftarrow$ $N$ users are split into $k$ groups, each with $N/k$ users\;
	\For{$i = 1,2,... ,k$}{
		$\hat l^*, \hat P \leftarrow $ run Algorithm~\ref{alg:practical_solution} with $\pi$\;
		\ForEach{$u \in U_i$}{
			\tcc{$u$ downloads $\hat P$ to the mobile client}
			$l_u \leftarrow$ a randomly selected frequent location\;
			$l_u^* \leftarrow$ obfuscating $l_u$ by $\hat P$\;
			\tcc{$u$ uploads $l_u^*$ to the server}
			$
			\pi_u'(l) \leftarrow \frac{\pi(l) \hat P(l_u^*|l)}{\sum_{l' \in \mathcal L} \pi(l') \hat P(l_u^*|l')}, \ \forall l \in \mathcal L
			$\;
			\label{alg_line:bayes}
		}
		$\pi \leftarrow$ the mean value of $\pi_u'$ over $u \in U_i$\;
	}
	$U^* \leftarrow \varnothing$\;
	\For{$j = k, k-1,... ,1$}{
		\label{alg_line:select_start}
		\ForEach{$u \in U_j$}{
			\If{$u$'s obfuscated location is $\hat l^*$}{
				$U^* \leftarrow U^* \cup \{u\}$\;
				\If{$|U^*| == \alpha$}{
					\Return $U^*$\;
				}
			}
		}
	}
	\label{alg_line:select_end}
	\Return $U^*$\;
	\label{alg_line:end}
	
	\caption{\footnotesize User selection with dynamic  estimating $\pi$.}
	\label{alg:practical_solution_with_estimate_pi}
\end{algorithm}

\subsubsection{Estimating Overall Location Distribution $\pi$.}

Previously, we assume that we have known the overall frequent location distribution $\pi$. This may be possible  when we have other sources to infer $\pi$, e.g., mobile call logs \cite{blondel2012data}. However, if we do not have such data, other methods are required to estimate $\pi$ along with user selection. We thus propose a Bayes rule based method to do user selection and $\pi$ estimation simultaneously, as shown in Algorithm~\ref{alg:practical_solution_with_estimate_pi}.

Our basic idea is using users' uploaded obfuscated locations to refine $\pi$. Note that our mechanism requires that each user uploads the obfuscated location only once to ensure differential privacy protection \cite{andres2013geo}. Hence, to preserve differential privacy, we split all the users into $k$ groups, get users' obfuscated locations group by group, and iteratively refine $\pi$ with the obfuscated locations from previous user groups. The key update formula of $\pi$ is the Bayes rule in line~\ref{alg_line:bayes}. In such a way, the estimated $\pi$ gradually reaches the actual $\pi$ after iterative refinements. As $\pi$ generally becomes more and more accurate, the final user selection is biased to the users in the groups who upload locations later (line~\ref{alg_line:select_start}-\ref{alg_line:select_end}). The number of groups $k$ balances the trade-off between algorithm running efficiency and solution quality --- larger $k$ updates $\pi$ more frequently, but costs more time as it involves $k$ iterations of running Algorithm~\ref{alg:practical_solution}.

Note that in real implementation, users who do not have any frequent locations can still upload `NULL' to the server. Then, we can estimate the percentage of users who can report locations from previous user groups. This can help us to set an appropriate $\beta$ used in the optimization so as to finally find $\alpha$ users with a probability of $\rho$.

\section{Experiments} 

In this section, we conduct empirical studies on three real user mobility datasets. We use Algorithm~\ref{alg:practical_solution_with_estimate_pi} for both single and multi-location coverage scenarios given its practicality (no need to foreknow $\pi$).

\subsection{Baselines}

\newenvironment{myitemize}
{ \begin{itemize}
		\setlength{\itemsep}{0pt}
		\setlength{\parskip}{0pt}
		\setlength{\parsep}{0pt}     }
	{ \end{itemize}                  } 

\begin{myitemize}
	\item \textit{Laplace}. The state-of-the-art method to achieve geographic differential privacy is based on the Laplace distribution~\cite{andres2013geo}.
	\item \textit{NO}. We use the No-Obfuscation (NO) policy, i.e., the users upload one of their real frequent locations to the server, to show an upper bound of the coverage. 
	\item \textit{Random}. We use the random user selection to serve as the lower bound of the coverage that can be achieved.
\end{myitemize}

\subsection{Datasets}

\begin{myitemize}
	\item \textit{FS} dataset~\cite{yang2016privcheck} contains 1083 Foursquare users' check-ins in New York, USA across near one year. We set the time period to a \textit{weekly} granularity, that is, the selected users are expected to visit the target locations in the next week. The studied area (Figure~\ref{fig:ny}) is split into 1km*1km grids. Among the 45 weeks of user mobility data, we use the last five weeks as the test time period, and first 40 weeks for mobility profiling.
	
	\item \textit{CMCC} dataset contains 1315 users' GPS trajectories in Hangzhou, China, for one month from one mobile operator. The time period is set to a \textit{daily} granularity. The studied area (Figure~\ref{fig:hz}) is split into 1km*1km grids. We use the first 18 weekdays for mobility profiling and the remaining four weekdays for testing.
	
	\item \textit{D4D} dataset~\cite{blondel2012data} includes 5378 users' two-week mobile phone call logs with cell tower locations in Abidjan, C\^ote d'Ivoire. The time period is set to a \textit{daily} granularity. The studied area (Figure~\ref{fig:abidjan}) is split into cell-tower-based regions~\cite{xiong2016icrowd,wang2017location}. We use the first nine weekdays for mobility profiling and the last one weekday for testing.
\end{myitemize}

\begin{figure}[t]
	\centering
	\begin{subfigure}[t]{.245\linewidth}
		\includegraphics[width=1\linewidth]{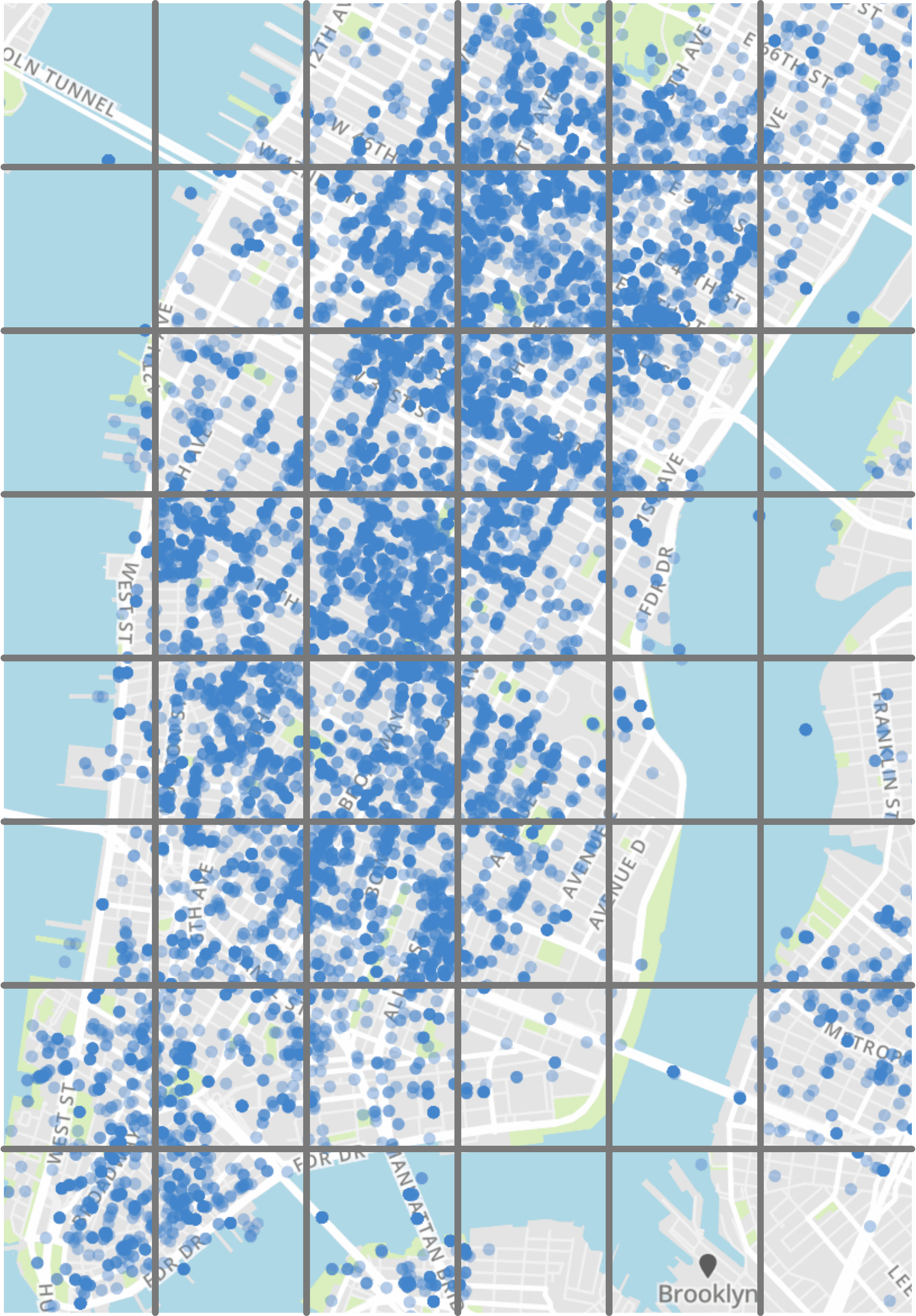}
		\caption{New York}
		\label{fig:ny}
	\end{subfigure}
	\begin{subfigure}[t]{.325\linewidth}
		\includegraphics[width=1\linewidth]{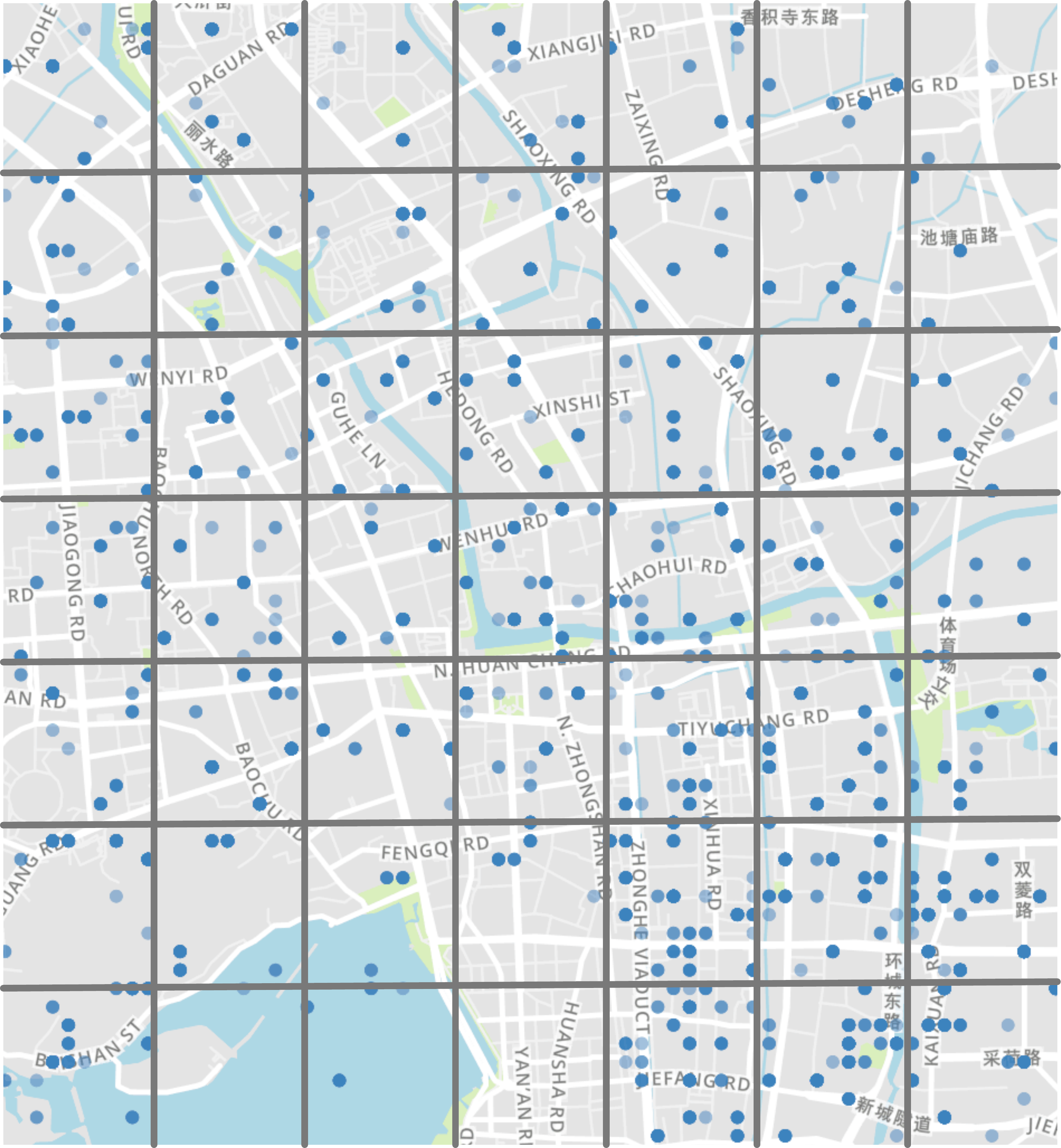}
		\caption{Hangzhou}
		\label{fig:hz}
	\end{subfigure}
	\begin{subfigure}[t]{.28\linewidth}
		\includegraphics[width=1\linewidth]{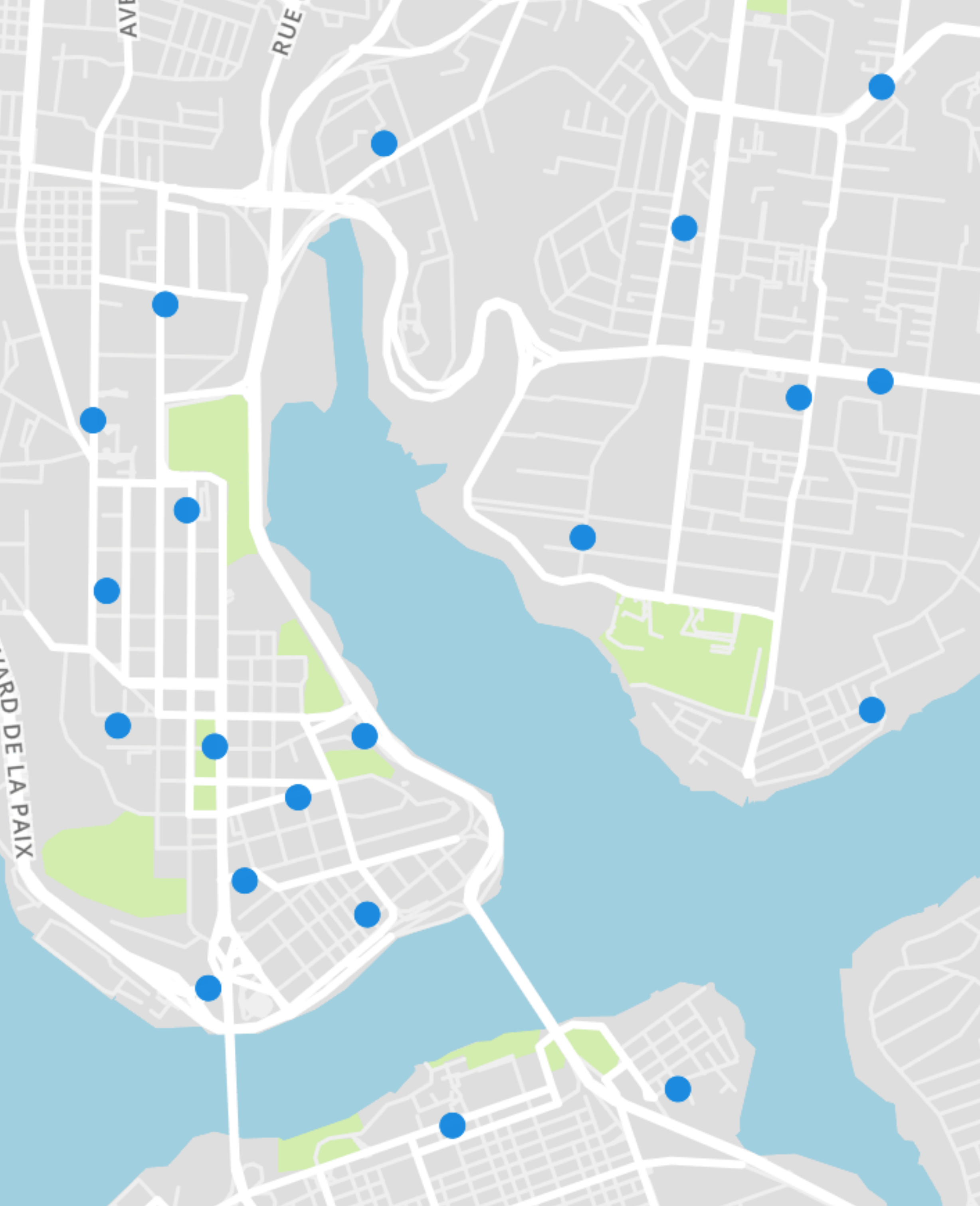}
		\caption{Abidjan}
		\label{fig:abidjan}
	\end{subfigure}
	\caption{Experiment areas. Points in NY and Hangzhou are user locations, and points in Abidjan are cell towers.}
\end{figure}

Table~\ref{tbl:parameters_exp} summarizes the experimental parameters. Note that the default differential privacy budget $\epsilon$ is set to $\ln(4)$ as suggested by the original paper~\cite{andres2013geo}.

\begin{table}[t]
	\renewcommand{\arraystretch}{1.2}
	\centering
	\scriptsize
	\caption{Experimental parameters.}
	\label{tbl:parameters_exp}
	\begin{tabular}{lll}
		\hline
		\textbf{Notation} & \textbf{Values} & \textbf{Description}                      \\
		\hline
		$\epsilon$           & ln(2), ln(4), ln(6), ln(8)          & differential privacy level              \\
		$\delta$               & 0.5, 0.6, 0.7, 0.8        & threshold for frequent locations     \\
		$N$ & 1083 (FS), 1315 (CMCC) & total number of users \\
		& 5378 (D4D) & \\
		$\alpha$            &  5\%$\cdot N$   & number of selected users \\
		$\rho$               & 95\% & probability for user selection \\
		$k$ & 6 & number of user groups \\
		\hline
	\end{tabular}
\end{table}

\subsection{Results on FS}

\subsubsection{Single Location Coverage.} 

We first evaluate the scenario where only one location (grid) needs to be covered. Our evaluation metric is the probability that a selected user will actually appear at the target location in the next week. 

\begin{figure}[t]
	\centering
	\begin{subfigure}[t]{0.26\linewidth}
		\includegraphics[width=1\linewidth]{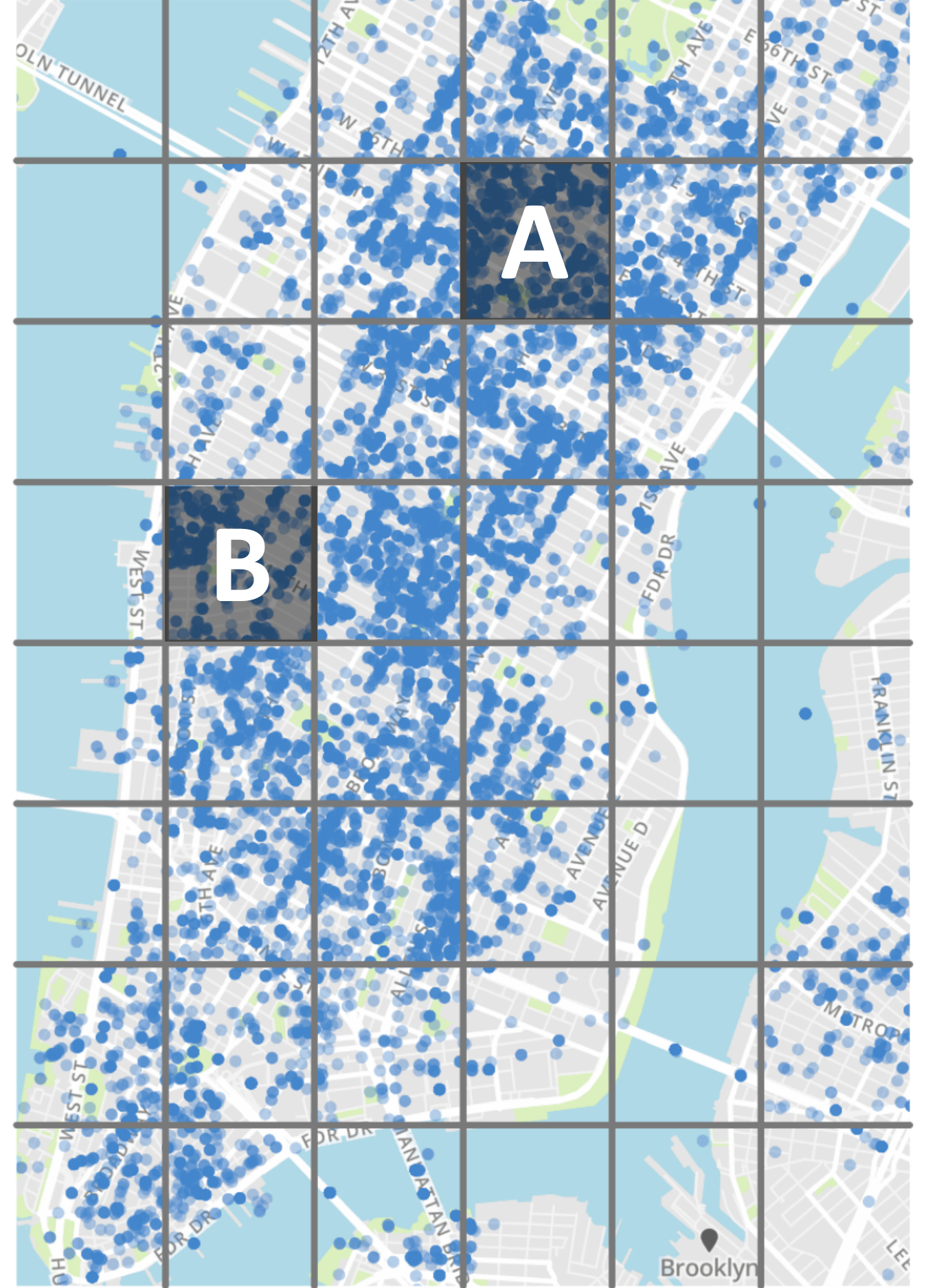}
	\end{subfigure}
	\begin{subfigure}[t]{0.62\linewidth}
		\includegraphics[width=1\linewidth]{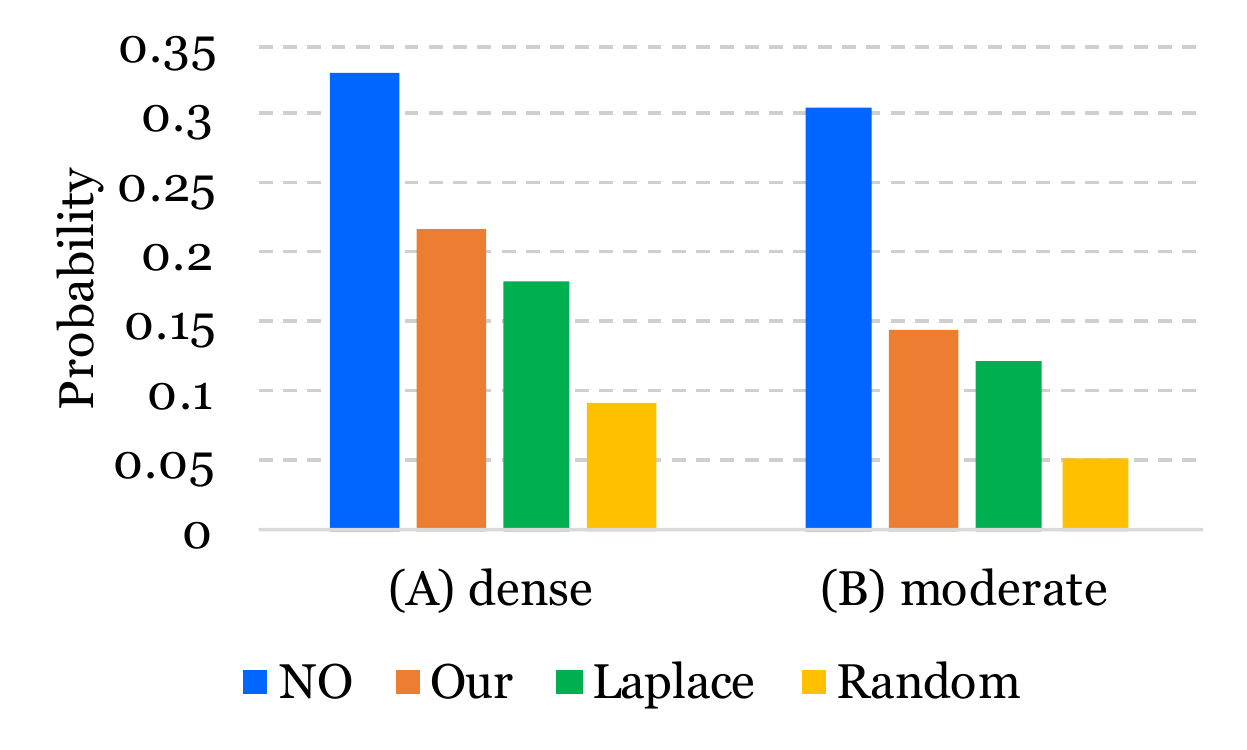}
	\end{subfigure}
	\caption{Experiment of single location coverage on two different populated locations on FS ($\epsilon = \ln(4), \delta=0.7$).}
	\label{fig:ny_single_week}
\end{figure}

Figure~\ref{fig:ny_single_week} shows the results on two target locations with different population sizes when $\epsilon=\ln(4)$ and $\delta=0.7$. In both target locations, our proposed method can achieve a larger coverage probability (up to 5\% improvement) than the Laplace mechanism. Compared to the no-obfuscation method, the coverage probability of our method drops from 32.9\% to 21.7\% for the densely populated target location. For the less densely populated one, the drop is bigger (from 30.5\% to 14.5\%). A possible explanation is that when the target location is densely populated, even if our mechanism mis-selects a user whose frequent location is not the target one, s/he still may go to the target location by chance.

\begin{figure}[t]
	\begin{subfigure}[t]{0.49\linewidth}
		\includegraphics[width=1\linewidth]{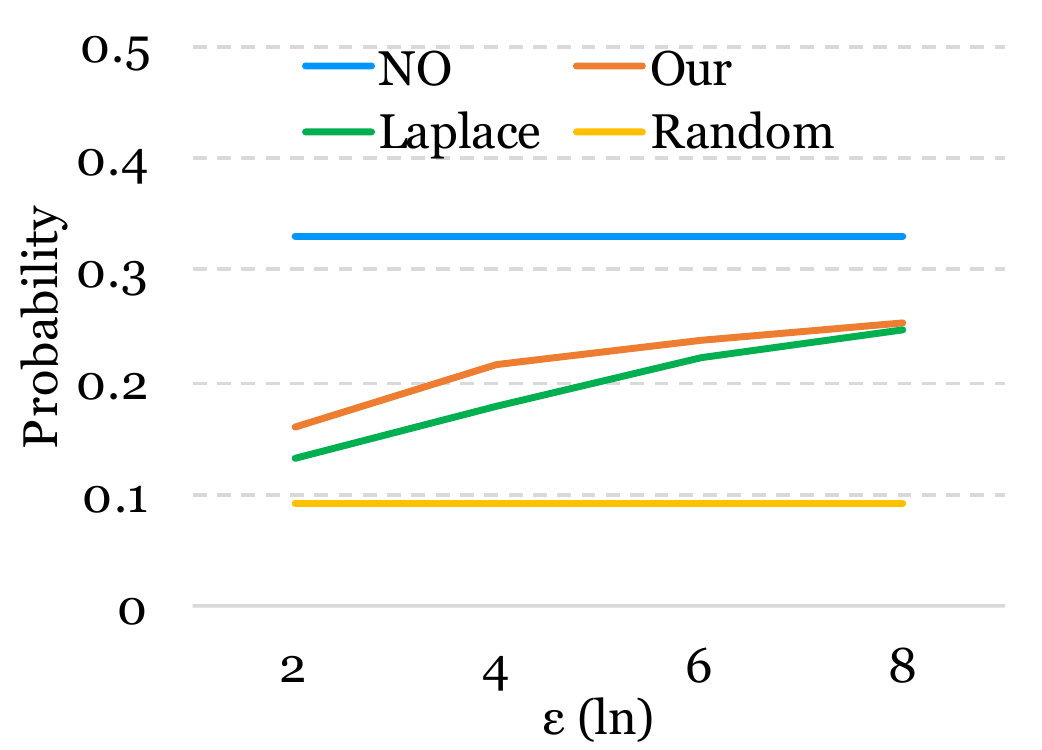}
		\caption{Varying $\epsilon$ ($\delta=0.7$)}
		\label{fig:ny_single_week_vary_epsilon}
	\end{subfigure}
	\begin{subfigure}[t]{0.49\linewidth}
		\includegraphics[width=1\linewidth]{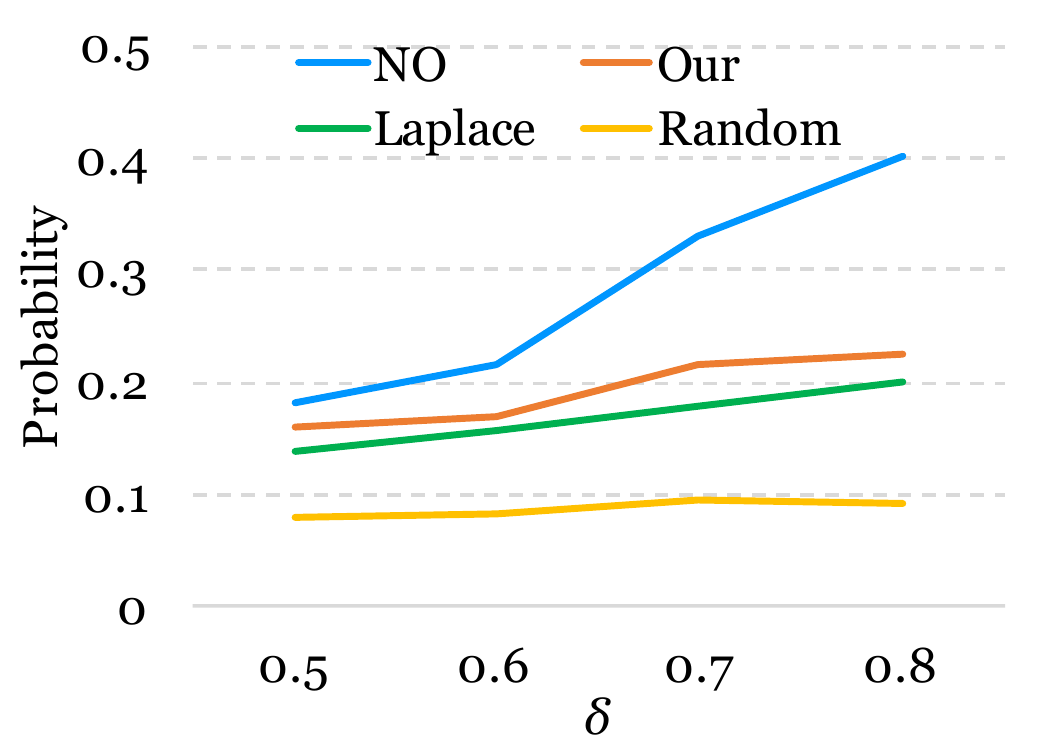}
		\caption{Varying $\delta$ ($\epsilon=\ln(4)$)}
		\label{fig:ny_single_week_vary_delta}
	\end{subfigure}
	\caption{Single location coverage results on FS.}
\end{figure}

Figure~\ref{fig:ny_single_week_vary_epsilon} illustrates how the coverage probability changes when we vary the privacy budget $\epsilon$ for the densely populated target location. As a trade-off between privacy and coverage, when $\epsilon$ increases (i.e., lower level of privacy), we can get a higher coverage probability. More specifically, the improvement of our method over Laplace is more significant for a lower $\epsilon$, i.e., higher privacy protection guarantee. 

Figure~\ref{fig:ny_single_week_vary_delta} shows the change of coverage probability when the threshold of frequent locations $\delta$ varies. The coverage probabilities of all the methods rise with the increase of $\delta$. While a higher $\delta$ benefits coverage probability, the number of users who can upload their (obfuscated) frequent locations (i.e., candidates for selection) is smaller, because only users with at least one location profiling probability larger than $\delta$ will upload frequent locations. Based on experiment results, setting $\delta$ to around 0.7-0.8 is appropriate for our method, as the coverage probability is relatively satisfactory while a large portion of users can be involved.

\begin{figure}[t]
	\centering
	\begin{minipage}{0.49\linewidth}
		\centering
		\includegraphics[width=1\textwidth]{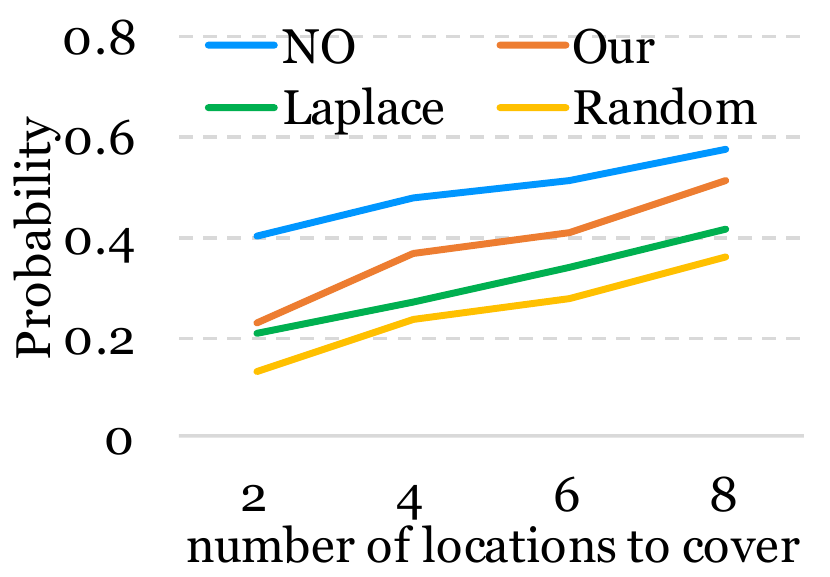}
		\caption{Multi-location coverage results on FS. 
		}
		\label{fig:ny_multi_loc}
	\end{minipage}\hfill
	\begin{minipage}{0.49\linewidth}
		\centering
		\includegraphics[width=1\textwidth]{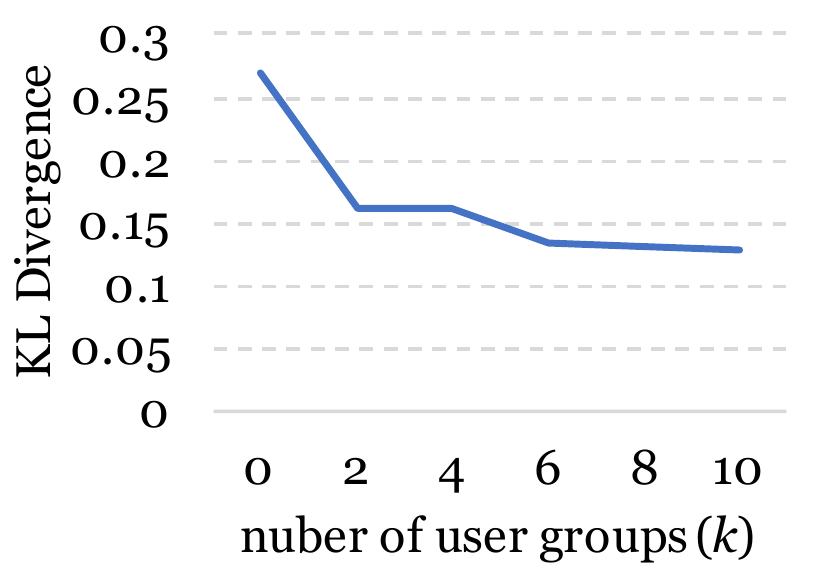}
		\caption{Results of estimating $\pi$ on FS.}
		\label{fig:kl}
	\end{minipage}
\end{figure}

\subsubsection{Multi-Location Coverage.} We evaluate the scenario where multiple target locations exist. We randomly select 2, 4, 6 and 8 locations as the targets. Figure~\ref{fig:ny_multi_loc} shows the actual coverage probability that we can get, i.e., the probabilities of selected users covering any one of the target locations in the coming week. The results show that our proposed method consistently outperforms Laplace under the same level of privacy protection. Moreover, with an increasing number of the target locations, we find that the performance gap between our method and no-obfuscation becomes smaller. This indicates that, when there are more locations to cover, using our mechanism is more profitable, as the performance loss incurred by the geographic differential privacy protection becomes smaller. 

\subsubsection{Estimation of $\pi$.} We evaluate whether our proposed Bayes rule based method can estimate $\pi$ correctly. We use KL divergence~\cite{kullback1951information} to quantify the similarity between the estimated $\pi$ and the actual $\pi$. 
The smaller KL divergence is, the more similar they are. Figure~\ref{fig:kl} shows the change of KL divergence with $k$ (the number of user groups), and $\pi$ is initialized to a uniform distribution. In Figure~\ref{fig:kl}, $k=0$ represents the KL divergence between the uniform and the actual distribution. When $k$ is small, we have fewer iterations to update $\pi$, leading to a larger KL divergence. In our experiment, $k=6$ is a good setting, as KL divergence achieves a relatively low value, while the algorithm can complete execution within a reasonable time.

\subsubsection{Runtime Efficiency.} We use Gurobi 7.5 \cite{optimization2014inc} as the linear programming solver engine to run Algorithm~\ref{alg:practical_solution} for getting the optimal policy $\hat P$. It takes about 450 seconds on a commodity laptop with i5-5200U (2.2 GHz), 8G memory. We split all the users to six groups, meaning that Algorithm~\ref{alg:practical_solution} is executed six times, which sums up to about 45 minutes. As the optimal privacy policy generation can be an offline process, such runtime efficiency is totally acceptable for real applications. Note that this running time  is not affected by the number of users, so our method can serve mobile applications with a large number of users.

\subsection{Results on CMCC and D4D}
\begin{figure}[t]
	\begin{subfigure}[t]{0.49\linewidth}
		\includegraphics[width=1\linewidth]{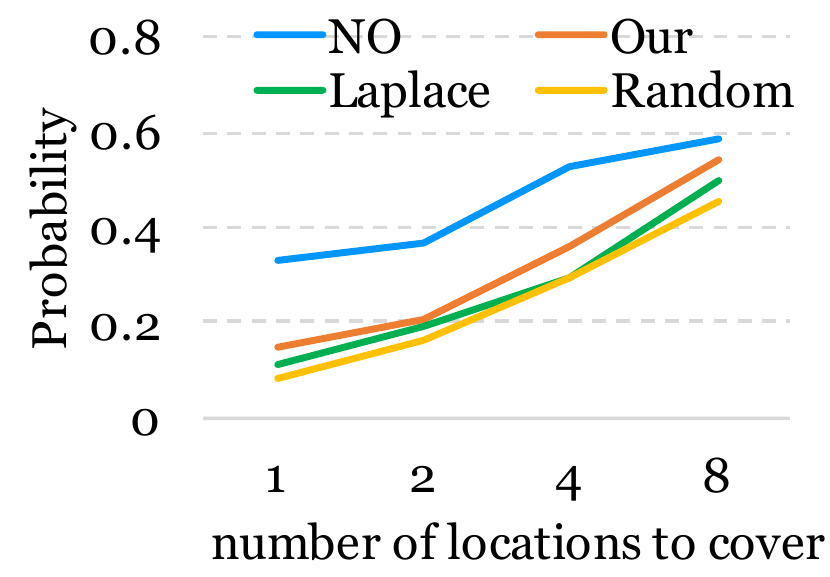}
		\caption{CMCC (Hangzhou)}
		\label{fig:cmcc_eva}
	\end{subfigure}
	\begin{subfigure}[t]{0.49\linewidth}
		\includegraphics[width=1\linewidth]{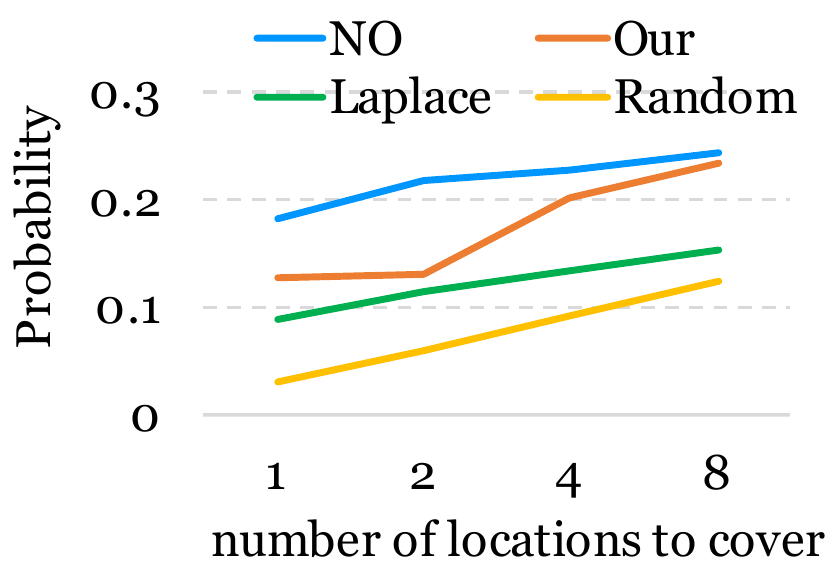}
		\caption{D4D (Abidjan)}
		\label{fig:d4d_eva}
	\end{subfigure}
	\caption{Experiment results on CMCC and D4D with different number of locations to cover ($\epsilon=\ln(4), \delta=0.8$).}
\end{figure}

To test the robustness of our proposed method, we also conduct experiments on CMCC and D4D datasets. The results are shown in Figure~\ref{fig:cmcc_eva} and~\ref{fig:d4d_eva}, where we randomly select 1, 2, 4, and 8 locations to cover. The results verify that our proposed method can always outperform the Laplace mechanism in attaining a higher coverage probability of the selected users. Moreover, the results show that when the number of target locations to cover increases to 8, our privacy mechanism almost achieves the same coverage probability as no-obfuscation, especially for the D4D dataset.
This further emphasizes the practicability of our mechanism, as user privacy is gained with a nearly negligible quality loss. Note that the achieved coverage probability on D4D is smaller than FS or CMCC in general, because the phone call locations on D4D are intrinsically more difficult to predict. Please refer to the appendix for detailed mobility prediction results.

\section{Related Work} 
\label{sec:related_work}

Selecting a set of users who can cover a set of locations in the near future is a very important problem for real applications like spatial crowdsourcing~\cite{chen2016spatial,zhang20144w1h} and location-based advertising~\cite{dhar2011challenges}. In most of previous research works, users' moving histories are known and hence their mobility patterns can be effectively modeled for predicting their future locations~\cite{xiong2016icrowd,guo2017activecrowd,yang2015modeling}. 

As user privacy is becoming more and more important nowadays, some pioneering works have started to model users' mobility or activity patterns based on privacy-preserving data. Geo-indistinguishability mechanisms are proposed for location-based query systems where users can submit their differentially obfuscated locations \cite{andres2013geo,bordenabe2014optimal}. PrivCheck is designed to enable personalized location-based advertising or recommendation with obfuscated user check-ins, so that users' sensitive information (e.g., age and gender) cannot be inferred by adversaries~\cite{yang2016privcheck}. In spatial crowdsourcing data acquisition, recent works also incorporate privacy mechanisms to protect participants' precise locations~\cite{wang2016differential,wang2017location,to2014framework,vergara2014privacy,pournajaf2014spatial}. While these studies have various applications, they usually focus on obfuscating users' \textit{current} locations. As far as we know, little previous work has studied the privacy-preserving future crowd coverage maximization problem based on users' obfuscated \textit{historical} mobility profiles which we specifically focus on in this paper.

\section{Conclusion} 

In this paper, we study the crowd coverage maximization problem under the privacy protection on user locations. The key idea is to select users who will probably visit certain locations in near future with their differentially obfuscated locations. 
To maximize the quality (coverage probability) of selected users under such a privacy protection scheme, an optimization problem is formulated to obtain the optimal privacy policy. We mathematically analyze the problem, and then propose a practical algorithm to obtain the optimal privacy policy.
Experiments on various real user mobility datasets have verified the effectiveness of our privacy mechanism. 
As future work, we plan to study the problem when a user can upload multiple obfuscated frequent locations. 

\section{Acknowledgment}
This research is partially supported by NSFC Grant no. 71601106, State Language Commission of China Key Program Grant no. ZDI135-18, Hong Kong ITF Grant no. ITS/391/15FX, the European Research Council (ERC) under the European Union’s Horizon 2020 research and innovation program (grant agreement 683253/GraphInt).

\bibliographystyle{aaai}
\bibliography{private_crowd_cover} 

\begin{thebibliography}{}

\bibitem[\protect\citeauthoryear{Andr{\'e}s \bgroup et al\mbox.\egroup
  }{2013}]{andres2013geo}
Andr{\'e}s, M.~E.; Bordenabe, N.~E.; Chatzikokolakis, K.; and Palamidessi, C.
\newblock 2013.
\newblock Geo-indistinguishability: Differential privacy for location-based
  systems.
\newblock In {\em Proc. CCS},  901--914.

\bibitem[\protect\citeauthoryear{Blondel \bgroup et al\mbox.\egroup
  }{2012}]{blondel2012data}
Blondel, V.~D.; Esch, M.; Chan, C.; Cl{\'e}rot, F.; Deville, P.; Huens, E.;
  Morlot, F.; Smoreda, Z.; and Ziemlicki, C.
\newblock 2012.
\newblock Data for development: the d4d challenge on mobile phone data.
\newblock {\em arXiv preprint arXiv:1210.0137}.

\bibitem[\protect\citeauthoryear{Bordenabe, Chatzikokolakis, and
  Palamidessi}{2014}]{bordenabe2014optimal}
Bordenabe, N.~E.; Chatzikokolakis, K.; and Palamidessi, C.
\newblock 2014.
\newblock Optimal geo-indistinguishable mechanisms for location privacy.
\newblock In {\em Proc. CCS},  251--262.

\bibitem[\protect\citeauthoryear{Boyd and Vandenberghe}{2004}]{boyd2004convex}
Boyd, S., and Vandenberghe, L.
\newblock 2004.
\newblock {\em Convex optimization}.
\newblock Cambridge university press.

\bibitem[\protect\citeauthoryear{Chen and Shahabi}{2016}]{chen2016spatial}
Chen, L., and Shahabi, C.
\newblock 2016.
\newblock Spatial crowdsourcing: Challenges and opportunities.
\newblock {\em IEEE Data Eng. Bull.} 39(4):14--25.

\bibitem[\protect\citeauthoryear{Cho, Myers, and
  Leskovec}{2011}]{cho2011friendship}
Cho, E.; Myers, S.~A.; and Leskovec, J.
\newblock 2011.
\newblock Friendship and mobility: user movement in location-based social
  networks.
\newblock In {\em Proc. KDD},  1082--1090.

\bibitem[\protect\citeauthoryear{Dhar and Varshney}{2011}]{dhar2011challenges}
Dhar, S., and Varshney, U.
\newblock 2011.
\newblock Challenges and business models for mobile location-based services and
  advertising.
\newblock {\em Communications of the ACM} 54(5):121--128.

\bibitem[\protect\citeauthoryear{Dwork}{2008}]{dwork2008differential}
Dwork, C.
\newblock 2008.
\newblock Differential privacy: A survey of results.
\newblock In {\em International Conference on Theory and Applications of Models
  of Computation},  1--19.

\bibitem[\protect\citeauthoryear{Fawcett}{2006}]{fawcett2006introduction}
Fawcett, T.
\newblock 2006.
\newblock An introduction to roc analysis.
\newblock {\em Pattern recognition letters} 27(8):861--874.

\bibitem[\protect\citeauthoryear{Guo \bgroup et al\mbox.\egroup
  }{2017}]{guo2017activecrowd}
Guo, B.; Liu, Y.; Wu, W.; Yu, Z.; and Han, Q.
\newblock 2017.
\newblock Activecrowd: A framework for optimized multitask allocation in mobile
  crowdsensing systems.
\newblock {\em IEEE Transactions on Human-Machine Systems} 47(3):392--403.

\bibitem[\protect\citeauthoryear{Gurobi}{2014}]{optimization2014inc}
Gurobi.
\newblock 2014.
\newblock Inc.,“gurobi optimizer reference manual,” 2014.
\newblock {\em URL: http://www. gurobi. com}.

\bibitem[\protect\citeauthoryear{Kullback and
  Leibler}{1951}]{kullback1951information}
Kullback, S., and Leibler, R.~A.
\newblock 1951.
\newblock On information and sufficiency.
\newblock {\em The annals of mathematical statistics} 22(1):79--86.

\bibitem[\protect\citeauthoryear{Pournajaf \bgroup et al\mbox.\egroup
  }{2014}]{pournajaf2014spatial}
Pournajaf, L.; Xiong, L.; Sunderam, V.; and Goryczka, S.
\newblock 2014.
\newblock Spatial task assignment for crowd sensing with cloaked locations.
\newblock In {\em Proc. MDM}, volume~1,  73--82.

\bibitem[\protect\citeauthoryear{Rossi \bgroup et al\mbox.\egroup
  }{2015}]{rossi2015privacy}
Rossi, L.; Williams, M.~J.; Stich, C.; and Musolesi, M.
\newblock 2015.
\newblock Privacy and the city: User identification and location semantics in
  location-based social networks.
\newblock In {\em Proc. ICWSM},  387--396.

\bibitem[\protect\citeauthoryear{To, Ghinita, and
  Shahabi}{2014}]{to2014framework}
To, H.; Ghinita, G.; and Shahabi, C.
\newblock 2014.
\newblock A framework for protecting worker location privacy in spatial
  crowdsourcing.
\newblock {\em Proc. of the VLDB Endowment} 7(10):919--930.

\bibitem[\protect\citeauthoryear{Vergara-Laurens, Mendez, and
  Labrador}{2014}]{vergara2014privacy}
Vergara-Laurens, I.~J.; Mendez, D.; and Labrador, M.~A.
\newblock 2014.
\newblock Privacy, quality of information, and energy consumption in
  participatory sensing systems.
\newblock In {\em Proc. PerCom},  199--207.

\bibitem[\protect\citeauthoryear{Wang \bgroup et al\mbox.\egroup
  }{2016}]{wang2016differential}
Wang, L.; Zhang, D.; Yang, D.; Lim, B.~Y.; and Ma, X.
\newblock 2016.
\newblock Differential location privacy for sparse mobile crowdsensing.
\newblock In {\em Proc. ICDM},  1257--1262.

\bibitem[\protect\citeauthoryear{Wang \bgroup et al\mbox.\egroup
  }{2017}]{wang2017location}
Wang, L.; Yang, D.; Han, X.; Wang, T.; Zhang, D.; and Ma, X.
\newblock 2017.
\newblock Location privacy-preserving task allocation for mobile crowdsensing
  with differential geo-obfuscation.
\newblock In {\em Proc. WWW},  627--636.

\bibitem[\protect\citeauthoryear{Xiong \bgroup et al\mbox.\egroup
  }{2016}]{xiong2016icrowd}
Xiong, H.; Zhang, D.; Chen, G.; Wang, L.; Gauthier, V.; and Barnes, L.~E.
\newblock 2016.
\newblock icrowd: Near-optimal task allocation for piggyback crowdsensing.
\newblock {\em IEEE Transactions on Mobile Computing} 15(8):2010--2022.

\bibitem[\protect\citeauthoryear{Yang \bgroup et al\mbox.\egroup
  }{2015}]{yang2015modeling}
Yang, D.; Zhang, D.; Zheng, V.~W.; and Yu, Z.
\newblock 2015.
\newblock Modeling user activity preference by leveraging user spatial temporal
  characteristics in lbsns.
\newblock {\em IEEE Transactions on Systems, Man, and Cybernetics: Systems}
  45(1):129--142.

\bibitem[\protect\citeauthoryear{Yang \bgroup et al\mbox.\egroup
  }{2016}]{yang2016privcheck}
Yang, D.; Zhang, D.; Qu, B.; and Cudr{\'e}-Mauroux, P.
\newblock 2016.
\newblock Privcheck: privacy-preserving check-in data publishing for
  personalized location based services.
\newblock In {\em Proc. UbiComp},  545--556.

\bibitem[\protect\citeauthoryear{Zhang \bgroup et al\mbox.\egroup
  }{2014}]{zhang20144w1h}
Zhang, D.; Wang, L.; Xiong, H.; and Guo, B.
\newblock 2014.
\newblock 4w1h in mobile crowd sensing.
\newblock {\em IEEE Communications Magazine} 52(8):42--48.

\bibitem[\protect\citeauthoryear{Zheng \bgroup et al\mbox.\egroup
  }{2014}]{zheng2014urban}
Zheng, Y.; Capra, L.; Wolfson, O.; and Yang, H.
\newblock 2014.
\newblock Urban computing: concepts, methodologies, and applications.
\newblock {\em ACM Transactions on Intelligent Systems and Technology} 5(3):38.

\end{thebibliography}

\section{Appendix}
\subsection{Detailed Proof of Lemma 1}
Suppose we have two different $\hat l^*$, i.e., $l_1^*, l_2^* \in \mathcal L$, and get two different optimal objective values
\begin{equation*}
\frac{\pi(l_t)\hat P_1(\hat l_1^*|l_t) }{\sum_{l \in \mathcal L} \pi(l) \hat P_1(\hat l_1^*|l)} < \frac{\pi(l_t)\hat P_2(\hat l_2^*|l_t) }{\sum_{l \in \mathcal L} \pi(l) \hat P_2(\hat l_2^*|l)}
\end{equation*}
 We now construct a new solution of $\hat P_1'$ when $\hat l^*=l_1^*$ as follows:
\begin{align*}
	&\hat P_1'(l_1^*|l) = \hat P_2(l_2^*|l), \ \forall l \in \mathcal L \\
	&\hat P_1'(l_2^*|l) = \hat P_2(l_1^*|l), \ \forall l \in \mathcal L \\
	&\hat P_1'(l^*|l) = \hat P_2(l^*|l), \ l^* \not = l_1^*, l^* \not = l_2^*, \forall l \in \mathcal L
\end{align*}
We can verify that all the constraints of the optimization still stand, and then $\hat P_1'$ is a feasible solution when $\hat l^* = l_1^*$, and then 
\begin{equation*}
 \frac{\pi(l_t)\hat P_1'(\hat l_1^*|l_t) }{\sum_{l \in \mathcal L} \pi(l) \hat P_1'(\hat l_1^*|l)} = \frac{\pi(l_t)\hat P_2(\hat l_2^*|l_t) }{\sum_{l \in \mathcal L} \pi(l) \hat P_2(\hat l_2^*|l)} > \frac{\pi(l_t)\hat P_1(\hat l_1^*|l_t) }{\sum_{l \in \mathcal L} \pi(l) \hat P_1(\hat l_1^*|l)}
\end{equation*}
This violates that $\hat P_1$ is the optimal solution when $\hat l^* = \hat l^*_1$. \qedsymbol

\subsection{Detailed Proof of Theorem 1}


With the following way to construct $\hat P$, 
\begin{align}
\hat P(\hat l^*|l) & = \theta e^{-\epsilon d(l, l_t)}, && \forall l \in \mathcal L 
\label{eq:optimal_p_1}
\\
\hat P(l^*|l) & = \frac{1-\theta e^{-\epsilon d(l, l_t)}}{|\mathcal L|-1}, && \forall  l^*, l \in \mathcal L \textit{ and } l^* \not = \hat l^*
\label{eq:optimal_p_2}
\end{align}
Then, for any $l \in \mathcal L$,
\begin{equation*}
\sum_{l^* \in \mathcal L} \hat P(l^*|l) = \theta e^{-\epsilon d(l, l_t)} + (|\mathcal L| - 1) \frac{1-\theta e^{-\epsilon d(l, l_t)}}{|\mathcal L|-1} = 1
\end{equation*}
So the probability sum constraint stands. We then prove that differential privacy constraint also stands. Note that \textit{Lemma 3} has proved that the differential privacy constraint holds if $l^* = \hat l^*$. Therefore, we only need to show that the differential privacy constraint also stands for $l^* \not = \hat l^*$. Next, we show how to select $\theta$ to ensure that this is true for any $l^* \not = \hat l^*$,
\begin{align}
	& \frac{\hat P(l^*|l_1)}{\hat P(l^*|l_2)} = \frac{1-\theta e^{-\epsilon d(l_1,l_t)}}{1-\theta e^{-\epsilon d(l_2, l_t)}} \le e^{\epsilon d(l_1, l_2)} \\
	\Rightarrow \quad &\theta \le \frac{e^{\epsilon d(l_1, l_2)}-1}{e^{-\epsilon (d(l_2,l_t)-d(l_1,l_2))} - e^{-\epsilon d(l_1,l_t)}}
	\label{eq:theta_equation}
\end{align}
It is worth noting that both the numerator and denominator in the right side of Eq.~\ref{eq:theta_equation} are larger than zero when $\epsilon > 0$. Hence, we can set $\theta$ to any positive value smaller than or equal to
\begin{equation}
\min_{l_1,l_2 \in \mathcal L} \frac{e^{\epsilon d(l_1, l_2)}-1}{e^{-\epsilon (d(l_2,l_t)-d(l_1,l_2))} - e^{-\epsilon d(l_1,l_t)}}
\end{equation}
and then for any $l^* \not = \hat l^*$, geographic differential privacy still holds. Then, based on Lemma 2, we can know that the $\hat P$ satisfying Eq.~\ref{eq:optimal_p_1} and \ref{eq:optimal_p_2} can lead to the upper bound of the objective value. \qedsymbol

\subsection{Detailed Proof of Lemma 5}

According to the geographic differential privacy constraints, we have
\begin{align}
	&e^{\epsilon d(l,l_t)} P(\hat l^*|l) \ge  P(\hat l^*|l_t), \ l \not \in \mathbb L \textit{ and } l_t \in \mathbb L 
	\label{eq:dp_lemma3_proof}
	\\
	\Rightarrow \ & \pi(l_t)e^{\epsilon d(l,l_t)} P(\hat l^*|l)\ge   \pi(l_t)P(\hat l^*|l_t), \ l \not \in \mathbb L \textit{ and } l_t \in \mathbb L \\
	\Rightarrow \ & \sum_{l_t \in \mathbb L}  \pi(l_t)e^{\epsilon d(l,l_t)} P(\hat l^*|l) \ge \sum_{l_t \in \mathbb L}  \pi(l_t)P(\hat l^*|l_t), \ l \not \in \mathbb L \\
	\Rightarrow \ &  P(\hat l^*|l) \ge \frac{\sum_{l_t \in \mathbb L} \pi(l_t)P(\hat l^*|l_t)}{\sum_{l_t \in \mathbb L}  \pi(l_t)e^{\epsilon d(l,l_t)}}, \ l \not \in \mathbb L
	\label{eq:p_multi_1}
\end{align}
Then,
\begin{align}
	& \frac{\sum_{l_t \in \mathbb L} \pi(l_t)P(l^*|l_t)}{\sum_{l \in \mathcal L} \pi(l) P(l^*|l)} \\
	= & \frac{\sum_{l_t \in \mathbb L} \pi(l_t)P(l^*|l_t)}{\sum_{l_t \in \mathbb L} \pi(l_t)P(l^*|l_t) +  \sum_{l \not \in \mathbb L} \pi(l) P(l^*|l)}
\end{align}
For the ease of presentation, we denote $C = \sum_{l_t \in \mathbb L} \pi(l_t)P(l^*|l_t)$,
\begin{align}
	& \frac{\sum_{l_t \in \mathbb L} \pi(l_t)P(l^*|l_t)}{\sum_{l \in \mathcal L} \pi(l) P(l^*|l)} \\
	= & \frac{C}{C +  \sum_{l \not \in \mathbb L} \pi(l) P(l^*|l)} \\
	\le & \frac{C}{C +  \sum_{l \not \in \mathbb L} \pi(l) \frac{C}{\sum_{l_t \in \mathbb L}  \pi(l_t)e^{\epsilon d(l,l_t)}}} \\
	= & (1+\sum_{l \not \in \mathbb L}  \sum_{l_t \in \mathbb L} \frac{\pi(l)}{ \pi(l_t)}e^{-\epsilon d(l,l_t)})^{-1}
	\label{eq:multi_obj_max}
\end{align}
\qedsymbol

\subsection{Mobility Profiling}

We consider two popular mobility profiling methods used in literature, and choose the better one in our experiments.

(1) \textbf{Frequency}~\cite{guo2017activecrowd}. This method counts daily (or weekly) frequency that a user visits a location in her/his historical mobility records. For example, suppose we have a user's 7-day mobility history and s/he visits a location $l_i$ in 5 days, then the daily visiting probability is 5/7.

(2) \textbf{Poisson}~\cite{xiong2016icrowd}. Given a user $u$'s average daily (or weekly) visiting times to location $l_i$ in the past, denoted as $\lambda_{u,i}$, then the Poisson process estimates that $u$ visits $l_i$ at least once in one day (week) is:
\begin{equation}
p_{u,i} = 1-e^{-\lambda_{u,i}}
\end{equation}

\begin{figure*}[t]
	\centering
	\begin{subfigure}[t]{0.32\linewidth}
		\includegraphics[width=1\linewidth]{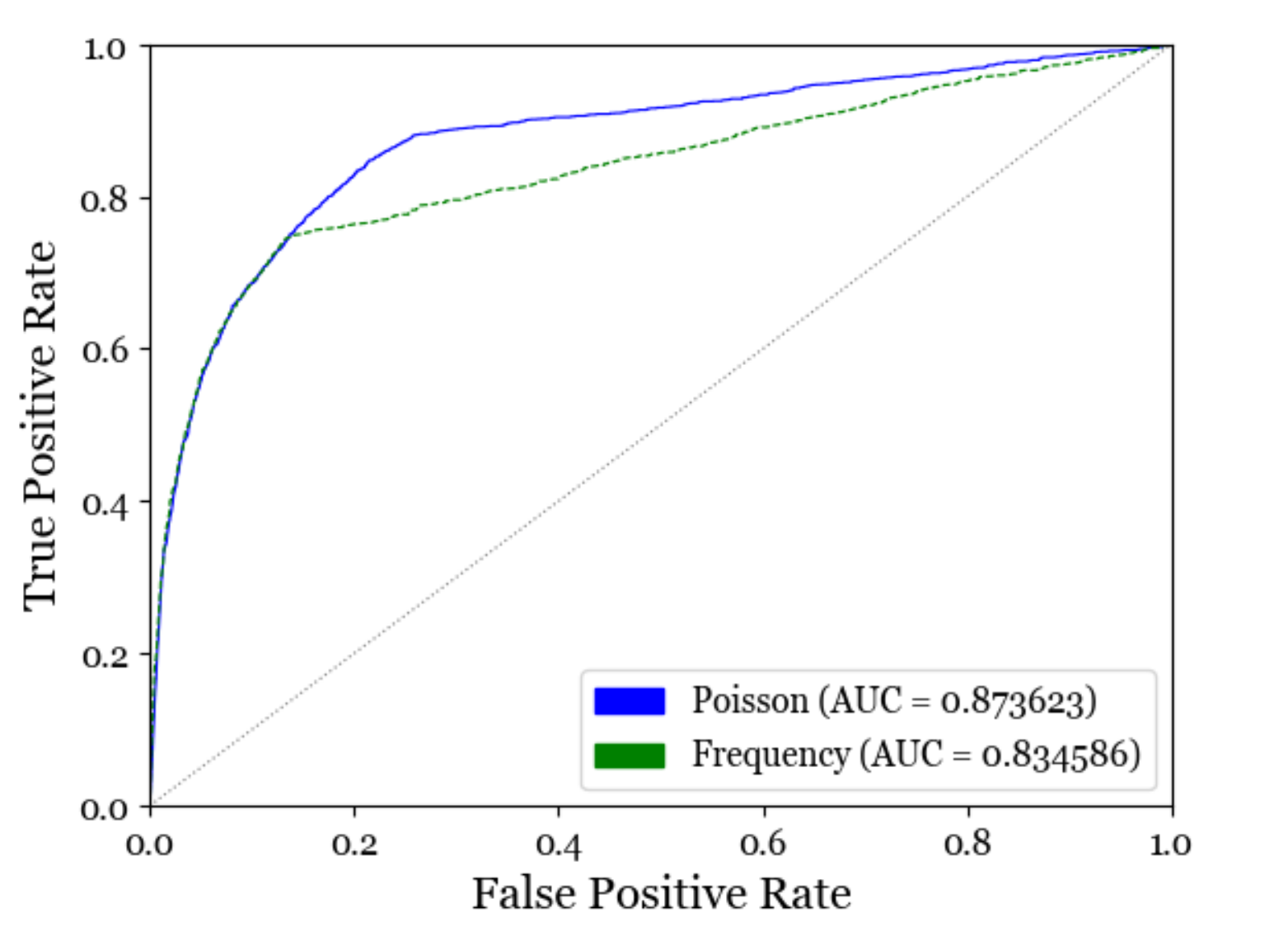}
		\caption{New York (FS)}
		\label{fig:ny}
	\end{subfigure}
	\begin{subfigure}[t]{0.33\linewidth}
		\includegraphics[width=1\linewidth]{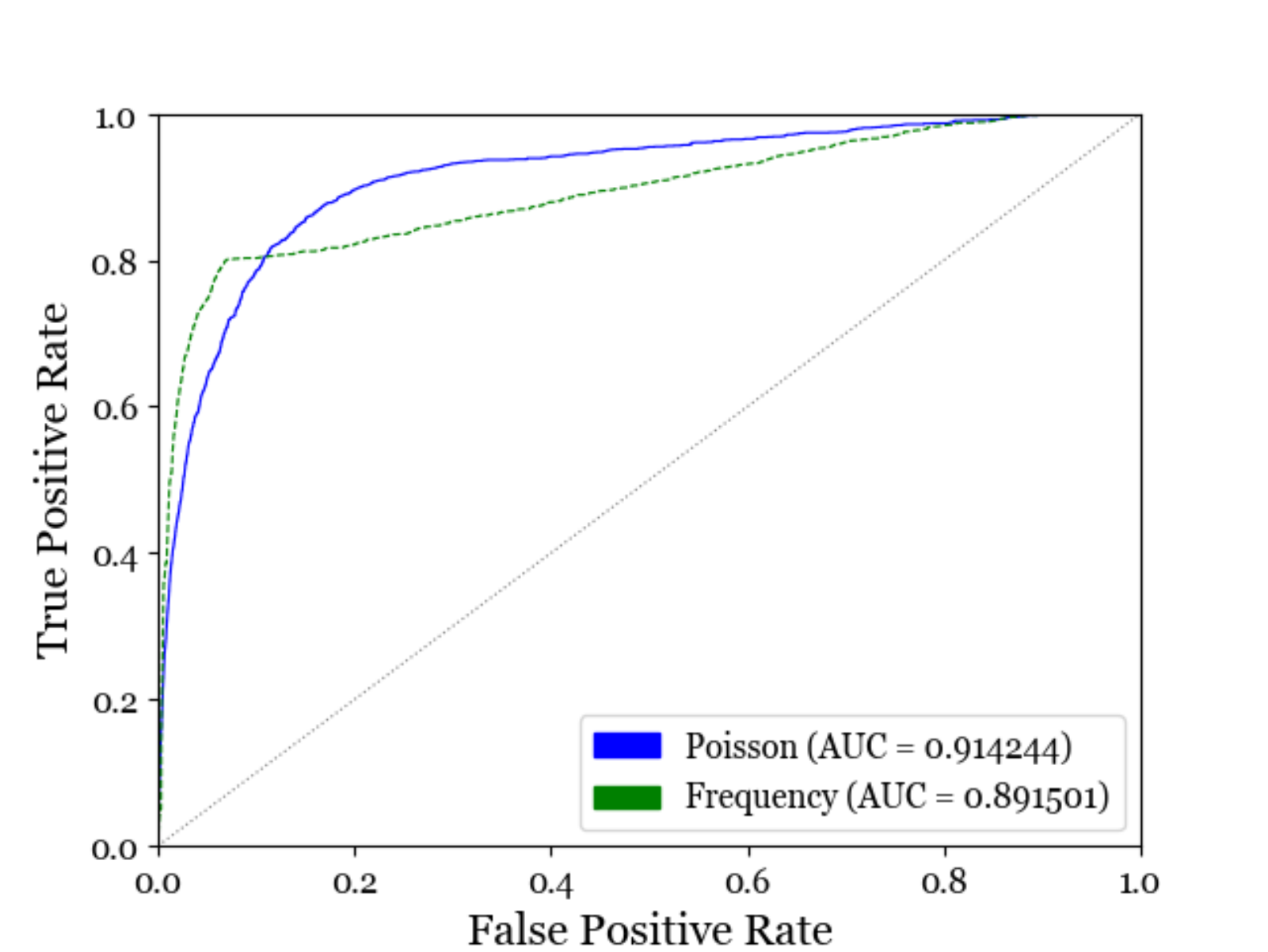}
		\caption{Hangzhou (CMCC)}
		\label{fig:hz}
	\end{subfigure}
	\begin{subfigure}[t]{0.32\linewidth}
		\includegraphics[width=1\linewidth]{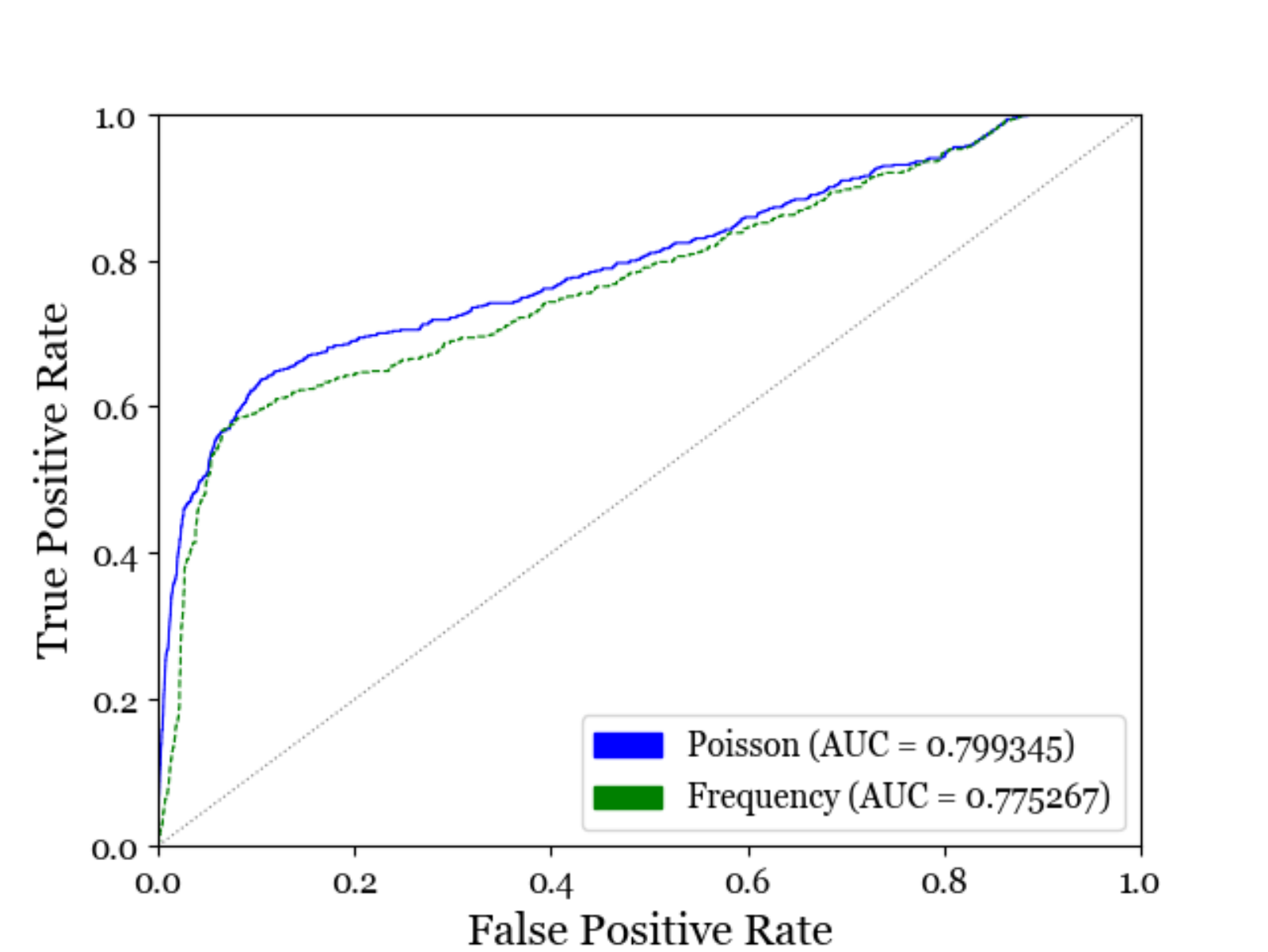}
		\caption{Abidjan (D4D)}
		\label{fig:abidjan}
	\end{subfigure}
	\vspace{-.5em}
	\caption{ROC curves of mobility profiling.}
	\label{fig:roc}
\end{figure*}

Figure~\ref{fig:roc} plots the receiver operating characteristics (ROC) curves~\cite{fawcett2006introduction} and shows the area under the ROC curve (AUC) values for both profiling methods on the FS, CMCC, and D4D datasets, respectively. The larger AUC value implies better performance in predicting a user's future mobility patterns. From the results, we see that Poisson-based mobility profiling method beats Frequency-based method, and thus we use the Poisson-based method in all the experiments. 

In addition, we observe that the mobility prediction on the D4D dataset is more difficult than on the other two datasets, as it gets a lower AUC value. As expected, our experiments in the paper (Figure~9) show that the selected users on the D4D dataset achieve a lower coverage probability than the other two datasets with the same user selection mechanism.

\end{document}